\documentclass[aps,pra,superscriptaddress,reprint,a4paper,longbibliography,nofootinbib]{revtex4-2}

\usepackage[utf8]{inputenc}
\usepackage[T1]{fontenc}
\usepackage{amsmath,amssymb,amsfonts}
\usepackage{mathtools}
\usepackage{graphicx}
\usepackage{array}
\usepackage{bm}
\usepackage{xcolor}
\usepackage{tikz} 
\usepackage{subfigure}
\usepackage{dsfont} 

\usepackage{float} 

\definecolor{myurlcolor}{rgb}{0,0,0.7}
\definecolor{myrefcolor}{rgb}{0.8,0,0}
\usepackage{hyperref}
\hypersetup{
    unicode=true, bookmarksnumbered=false, bookmarksopen=false, breaklinks=false, pdfborder={0 0 0},
    colorlinks=true,
    linkcolor=myrefcolor,citecolor=myurlcolor,urlcolor=myurlcolor
}

\graphicspath{ {./figs/} }
\makeatletter

\newcommand{\mcl}[1]{\mathcal{#1}}
\newcommand{\mbf}[1]{\mathbf{#1}}
\newcommand{\mrm}[1]{\mathrm{#1}}
\newcommand{\tbf}[1]{\textbf{#1}}
\newcommand{\trm}[1]{\textrm{#1}}

\newcommand{\dd}{\mathrm{d}}
\newcommand{\ee}{\mathrm{e}}
\newcommand{\ii}{\mathrm{i}}
\newcommand{\id}{\openone} 

\newcommand{\TT}{\mrm{T}}

\newcommand{\trace}{\mathrm{Tr}}
\usepackage{xparse}
\NewDocumentCommand\tr{g}{
  \IfNoValueTF{#1}
    {\trace}
    {\trace\!\left\{#1\right\}}
}

\newcommand{\imag}{\mrm{Im}}
\newcommand{\ket}[1]{\left| #1 \right\rangle }
\newcommand{\bra}[1]{\left\langle #1 \right| }
\newcommand{\ketbra}[2]{\ket {#1}\!\! \bra {#2} }

\newcommand{\eref}[1]{(\ref{#1})}
\newcommand{\eqnref}[1]{Eq.~(\ref{#1})}
\newcommand{\eqnsref}[2]{Eqs.~(\ref{#1}) and (\ref{#2})}
\newcommand{\figref}[1]{Fig.~\ref{#1}}
\newcommand{\tabref}[1]{Table~\ref{#1}}
\newcommand{\secref}[1]{Sec.~\ref{#1}}
\newcommand{\appref}[1]{App.~\ref{#1}}

\newcommand{\citeref}[1]{Ref.~\cite{#1}}
\newcommand{\refcite}[1]{Ref.~\cite{#1}}

\newcommand{\mVrms}{\mathrm{mV_{rms}}}
\newcommand{\ww}{\omega}
\newcommand{\gam}{\gamma}
\newcommand{\gamt}{\widetilde{\gamma}}
\newcommand{\GG}{\Gamma}

\newcommand{\Ttens}{{{\mcl T}}}

\renewcommand{\ss}{{\mrm{ss}}} 
\newcommand{\gmr}{\gamma_\mrm{gmr}} 

\newcommand{\rmdn}[2]{#1_{\mrm{#2}}}

\newcommand{\mean}[1]{\left\langle #1 \right\rangle}
\newcommand{\mvec}[1]{\tbf{#1}} 
\newcommand{\mmat}[1]{\tbf{#1}}
\renewcommand{\vec}[1]{\bm{#1}} 

\newcommand{\noise}{noise} 
\newcommand{\Hrf}{\hat{H}_{\mrm{\noise}}}
\newcommand{\Brf}{B_{\mrm{\noise}}}
\newcommand{\BrfVec}{\vec{B}_{\mrm{\noise}}}
\newcommand{\Omegarf}{\rmdn{\Omega}{\noise}}
\newcommand{\omegarf}{\rmdn{\omega}{\noise}}
\newcommand{\frf}{\rmdn{f}{\noise}}
\newcommand{\Vrf}{\rmdn{V}{\noise}}

\definecolor{lime}{HTML}{A6CE39}
\DeclareRobustCommand{\orcidicon}{
	\begin{tikzpicture}
	\draw[lime, fill=lime] (0,0) 
	circle [radius=0.16] 
	node[white] {{\fontfamily{qag}\selectfont \tiny ID}};
	\draw[white, fill=white] (-0.0625,0.095) 
	circle [radius=0.007];
	\end{tikzpicture}
	\hspace{-2mm}
}
\foreach \x in {A, ..., Z}{\expandafter\xdef\csname orcid\x\endcsname{\noexpand\href{https://orcid.org/\csname orcidauthor\x\endcsname}
			{\noexpand\orcidicon}}
}

\makeatother

\begin{document}

\title{Spin noise spectroscopy of an alignment-based atomic magnetometer}

\author{M.~Ko\'{z}bia\l\orcidA{}}
\altaffiliation{These authors contributed equally to this work.}
\affiliation{Centre for Quantum Optical Technologies, Centre of New Technologies, University of Warsaw, Banacha 2c, 02-097 Warszawa, Poland}

\author{L.~Elson\orcidB{}}
\altaffiliation{These authors contributed equally to this work.}
\affiliation{School of Physics and Astronomy, University of Nottingham, University Park, Nottingham, NG7 2RD, UK}

\author{L.~M.~Rushton\orcidC{}}
\affiliation{School of Physics and Astronomy, University of Nottingham, University Park, Nottingham, NG7 2RD, UK}

\author{\\A.~Akbar\orcidD{}}
\affiliation{School of Physics and Astronomy, University of Nottingham, University Park, Nottingham, NG7 2RD, UK}

\author{A.~Meraki\orcidE{}}
\affiliation{School of Physics and Astronomy, University of Nottingham, University Park, Nottingham, NG7 2RD, UK}

\author{K.~Jensen\orcidG{}} 
\email{kasjensendk@gmail.com}
\affiliation{School of Physics and Astronomy, University of Nottingham, University Park, Nottingham, NG7 2RD, UK}

\author{J.~Ko\l{}ody\'{n}ski\orcidF{}}
\email{jan.kolodynski@cent.uw.edu.pl}
\affiliation{Centre for Quantum Optical Technologies, Centre of New Technologies, University of Warsaw, Banacha 2c, 02-097 Warszawa, Poland}


\begin{abstract}
Optically pumped magnetometers (OPMs) are revolutionising the task of magnetic-field sensing due to their extremely high sensitivity combined with technological improvements in miniaturisation which have led to compact and portable devices. OPMs can be based on spin-oriented or spin-aligned atomic ensembles which are spin-polarised through optical pumping with circular or linear polarised light, respectively. Characterisation of OPMs and the dynamical properties of their noise is important for applications in real-time sensing tasks.
In our work, we experimentally perform spin noise spectroscopy of an alignment-based magnetometer. Moreover, we propose a stochastic model that predicts the noise power spectra exhibited by the device when, apart from the strong magnetic field responsible for the Larmor precession of the spin, white noise is applied in the perpendicular direction aligned with the pumping-probing beam. By varying the strength of the noise applied as well as the linear-polarisation angle of incoming light, we verify the model to accurately predict the heights of the Larmor-induced spectral peaks and their corresponding linewidths. Our work paves the way for alignment-based magnetometers to become operational in real-time sensing tasks.
\end{abstract}

\maketitle

\section{Introduction}
Optically pumped magnetometers (OPMs)~\cite{budker_romalis_2007, BudkerJacksonKimball2013} based on e.g.~caesium, rubidium or potassium atomic vapour, or helium gas can have high sensitivity in the fT/$\sqrt{\mrm{Hz}}$ range~\cite{kominis_kornack_allred_romalis_2003}. Commercially available OPMs include scalar OPMs for use in e.g.~geophysical surveys~\cite{gemsys,quspin} and zero-field OPMs~\cite{quspin, twinleaf,fieldline, m4h} which are promising for applications within areas such as~cardiology~\cite{Bison2009apl, Wyllie2012ol, Jensen2018scirep} and neuroscience~\cite{Xia2006, Boto2018nature, LabytSanderWakai2022}. OPMs can also be used for detection of radio-frequency (RF) magnetic fields with potential applications within biomedical imaging~\cite{marmugi_renzoni_2016, Jensen2019}, non-destructive testing~\cite{Wickenbrock2016apl, Bevington2019jap}, and remote sensing~\cite{deans_marmugi_renzoni_2018, rushton_2022}. However, such RF OPMs are not yet commercially available. Orientation-based RF OPMs are typically implemented using two or three laser beams~\cite{wasilewski_2010,Chalupczak2012,Keder2014}. On the other hand, alignment-based RF OPMs implemented with a single laser beam~\cite{ledbetter2007detection} are promising for applications and commercialisation~\cite{rushton2023alignment}.

In an optical magnetometer, the atoms are spin-polarised using light through the process of optical pumping~\cite{Happer1972}. In an orientation-based optical magnetometer, the atoms are optically pumped with circularly polarised light. In this case, each atom can be effectively treated as a spin-1/2 particle, even if the ground state of the atom has a total angular momentum $F$ larger than 1/2. The evolution of the atomic spin in a magnetic field is then well described by the Bloch equation for the spin vector $\bm{F}=\left(F_x,F_y,F_z \right)^\TT $. Its three components correspond to the expectation values of the angular momentum operators defined along the respective directions, i.e.~$F_\alpha \coloneqq \tr\{\hat{F}_\alpha\rho\}$ with $\alpha=x,y,z$ for an atomic ensemble being effectively described by a single-atom density matrix $\rho$. On the contrary, in an alignment-based magnetometer~\cite{weis2006theory, ledbetter2007detection, zigdon2010nonlinear, Beato2018pra, Bertrand2021rsi, rushton2023alignment} the atoms are optically pumped with linearly polarised light. In this case, each atom can be effectively treated as a spin-1 particle~\cite{Colangelo2013}. As a result, one has to abandon describing the atomic state with a three-component vector, and instead describe it using rank-2 spherical tensors with five components, which describe how the atomic spin is aligned along certain axes~\cite{auzinsh_budker_rochester_2014}.

The purpose of \emph{spin noise spectroscopy} (SNS)~\cite{sinitsyn2016theory} is to characterise the noise properties of a given atomic system and, in particular, the form of the autocorrelation function that noise  fluctuations exhibit in the steady-state regime~\cite{GardinerZoller}. However, only in the case of orientation-based magnetometers have stochastic noise models been proposed that are capable of explaining the observed noise power spectra when probing unpolarised atomic ensembles~\cite{Zapasskii2013,Glasenapp2014,Shah2010,lucivero2016squeezed}, also including the effects of spin-exchange collisions~\cite{Wen2021,Mouloudakis2022,Mouloudakis2023}. In contrast, such models characterising fully the spin-noise spectra in alignment-based magnetometers are still missing, despite recent promising steps in that direction~\cite{fomin2020spin,Fomin2021,Liu2022,Liu2023,Delpy2023,Delpy2023njp}.

In our work, we employ methods of stochastic calculus and the formalism of spherical tensors to predict the \emph{power spectral density} (PSD) of an alignment-based magnetometer in the presence of a strong static magnetic field affected by white noise that is applied in the perpendicular direction, i.e.~along the beam simultaneously pumping and probing the ensemble%
\footnote{
	In contrast to the parallel configuration, in which the white noise would just yield effective fluctuations of the static field~\cite{Delpy2023njp}.
}. Our model correctly predicts the existence of peaks in the measured PSD at particular multiples of the Larmor frequency, as well as the dependence of their amplitudes and widths on the system geometry and the noise intensity. Importantly, we verify our model in a series of experiments, whose results show very good accordance with the predictions.

Our work paves the way for exploring alignment-based magnetometers in real-time sensing tasks, in which---thanks to the detailed characterisation of the spin noise---one is capable of tracking time-varying signals beyond the nominal bandwidth dictated by the magnetometer~\cite{Jimenez2018}. Our alignment-based magnetometer with added noise can be used as a scalar magnetometer to sense time-varying magnetic fields, but also potentially ones that oscillate at RF. Thanks to employing only a single beam of light for both pumping and probing the atoms, the simplicity of the proposed architecture is promising with respect to potential miniaturisation and commercialisation~\cite{Kitching2018}.

The remainder of the paper is organised as follows. We firstly describe the spatial configuration of the magnetometer considered in \secref{sec:magnetometer}, in order to motivate and explain the spherical-tensor formalism that we particularly employ to parametrise its atomic state in \secref{sec:steady_state}. In \secref{sec:dynamics} we then discuss the evolution of the atomic state and, in particular, how it determines the dynamics of relevant spherical-tensor components and the detected signal, so that in \secref{sec:setup} we may relate it to and describe in detail the physical parameters of our experimental setup. In \secref{sec:SNS} we turn to SNS that constitutes the goal of our work. We firstly explain in \secref{sec:predictions} what the form of PSD is that we expect for the magnetometer considered, and how we predict it. The results of the experiment are then shown in \secref{sec:results} and compared to the theory in \secref{sec:validation} and \secref{sec:validation2}. Finally, we conclude in \secref{sec:conclusions}.

\section{Alignment-based atomic magnetometer}
\label{sec:magnetometer}
\begin{figure}[!t]
\includegraphics[width=\columnwidth]{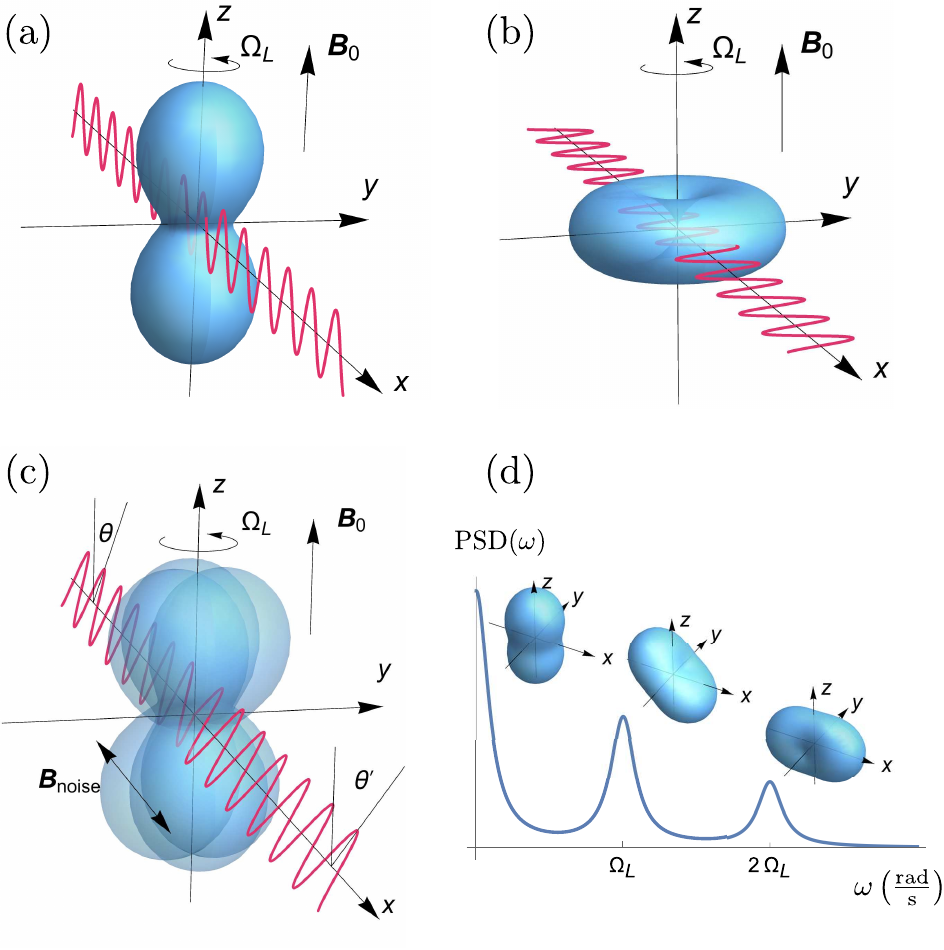}
\caption{\textbf{Alignment-based magnetometer:~spatial configuration}. The light propagates in the direction $x$ in the presence of a strong, static magnetic field $\vec{B}_0$, pointing along the $z$-direction. In (a) and (b), angular-momentum probability surfaces of the resulting atomic steady-state are presented, when the atom is pumped by vertically ($\theta=0$) or horizontally ($\theta=\pi/2$) polarised light, respectively. As shown in (c), considering the input light-beam to be polarised at some intermediate angle $0<\theta<\pi/2$, the atomic steady-state generally modifies the polarisation angle of the transmitted light $\theta'$. Moreover, in our experiment a stochastic field $\vec{B}_\mrm{\noise}$ is applied along the direction of light propagation, and induces white noise that disturbs the atomic state from equilibrium. The so-created randomly tilted state has now three distinct contributions from spherical components $\Ttens^{(2)}_{0}$, $\Ttens^{(2)}_{\pm 1}$ and $\Ttens^{(2)}_{\pm 2}$, whose angular-momentum probability surfaces shown in (d) exhibit periodicity under rotations around $z$ at multiples $0$, $\Omega_L$ and $2\Omega_L$ of the Larmor (angular) frequency $\Omega_L$, respectively. As a result, as depicted in (d), the recorded \emph{power spectral density} (PSD) of $\theta'$-fluctuations should contain three peaks at these particular frequencies with amplitudes dictated by the contribution of each spherical-tensor component.
}
\label{fig:geometry}
\end{figure}
In \figref{fig:geometry} we depict the natural spatial configuration of an alignment-based atomic magnetometer~\cite{ledbetter2007detection}, which applies to the implementation considered here. A strong magnetic field $\vec{B}_0$, directed along $z$, acts perpendicularly to the direction of light propagation, here chosen to be $x$, with light being linearly polarised at an angle $\theta$ to $z$-axis in the $yz$-plane. Both the field and the light interact with an atom and modify its steady state, otherwise induced solely by the relaxation processes~\cite{weis2006theory}. Although we defer the formal description of the atomic steady-state, let us note for now that its form can be conveniently represented by the \emph{angular-momentum probability surface}~\cite{Rochester2001ajp, auzinsh_budker_rochester_2014} (depicted in blue in \figref{fig:geometry}). As the surface describes the effective polarisation of the steady state, it indicates the rotational symmetries the state possesses. Its shape strongly depends on the polarisation angle $\theta$ of the input light-beam, e.g., resembling a ``peanut'' for $\theta=0$ or a ``doughnut'' for $\theta=\tfrac{\pi}{2}$, see \figref{fig:geometry}(a) and (b), respectively. As such a qualitative description, however, relies solely on symmetry arguments, it applies irrespectively of other properties of the input light, e.g., its detuning.

Importantly, in our experiment the atom is further disturbed by a stochastic field $\rmdn{\vec{B}}{\noise}$, see \figref{fig:geometry}(c), that induces white noise along the $x$-direction, whose impact can then be monitored by inspecting the polarisation angle $\theta'$ of the transmitted beam, while the atomic state is constantly ``kicked out'' of equilibrium. As we carry out the experiment in conditions in which correlations between distinct atoms may be neglected during the sensing process, the dynamics of the whole ensemble is effectively captured by the evolution of a single atom~\cite{Happer1972}. Furthermore, due to the effective atomic density being low, the shot noise of the photon-detection process dominates over the atomic projection noise, and the impact of the measurement backaction on the atomic steady-state can be ignored~\cite{Deutsch2010,Troullinou2021}. Moreover, thanks to the presence of a relatively strong magnetic field $\vec{B}_0$, given that the experiment is carried out at room temperature, the mechanism of spin-exchange interactions during atomic collisions is irrelevant~\cite{Savukov2005}, in contrast to SERF magnetometers~\cite{Savukov2017}, that could in principle induce not only classical correlations but also entanglement between individual atoms~\cite{Mouloudakis2021,Katz2022}. 

In what follows, we firstly formalise the description of an atomic steady-state in the spherical-tensor representation, in order to then show how such a formalism allows us to compactly model the magnetometer dynamics, as well as predict the behaviour of the detected signal.

\subsection{Spherical tensor representation of the atomic steady state} 
\label{sec:steady_state}
Any density matrix describing an atom of total angular momentum $F$ can always be written in the eigenbasis of the angular-momentum operators $\hat{F}^2$ and $\hat{F}_z$ as~\cite{Happer1972}:
\begin{equation}
    \rho^{(F)}=\sum_{M,M'=-F}^{F}\rho_{MM'}^{(F)}\ketbra {F,M\phantom{'}}{F,M'}.
\label{eq:rho_FM}
\end{equation}
However, it is much more convenient to decompose the density matrix in the basis of \emph{spherical tensor operators} of rank $\kappa=0,1,\dots,2F$, which transform independently in each $\kappa$-subspace under rotations, and hence magnetic fields, in a well-behaved manner. In particular, the atomic state of fixed $F$ in \eqnref{eq:rho_FM} can be generally written as:
\begin{equation}
    \rho^{(F)}=\sum_{\kappa=0}^{2F}\sum_{q=-\kappa}^{\kappa}m_{\kappa q}^{(F)}\,\Ttens^{(\kappa)}_q\!(F),
    \label{eq:rho_m}
\end{equation}
where the density matrix is now decomposed into a sum of rank-$\kappa$ \emph{components}, each constituting a sum (over $q$) of, in principle, non-Hermitian \emph{tensor operators} $\Ttens^{(\kappa)}_q\!(F)$ (see e.g.~\cite{Happer1972} and \appref{app:spherical_tensors}) multiplied by their corresponding complex \emph{coefficients} $m_{\kappa q}^{(F)}$. For a given fixed $F$, $\Ttens^{(\kappa)}_q$ form a basis and they formally satisfy $\tr{\Ttens^{(\kappa)}_q (\Ttens^{(\kappa')}_{q'})^\dagger}=\delta_{\kappa\kappa'}\delta_{qq'}$, while it is convenient to also impose $\Ttens^{(\kappa)\dag}_q=(-1)^q\Ttens^{(\kappa)}_{-q}$, so that conditions $m_{\kappa q}=(-1)^q m^{*}_{\kappa,-q}$ and $\imag \{m_{\kappa0}\}=0$ ensure then the density matrix \eqref{eq:rho_m} is Hermitian. The $m_{00}=1/\sqrt{2F+1}$ coefficient is fixed by the $\tr{\rho}=1$ condition, while the corresponding ($\kappa=0$) tensor operator, $\Ttens^{(0)}_0=\tfrac{1}{\sqrt{2F+1}}\openone_{2F+1}$, is the only one with a non-zero trace, being invariant under any rotations.

In principle, the atomic steady state may involve more than one $F$-level, e.g.~$F=3,4$ in the case of the D1-line transition in caesium. However, if the laser field is tuned to a specific optical transition from a single $F$-level, and there is no coherent coupling between levels of different $F$, one can disregard coherences between these and most generally write the atomic steady state as:
\begin{equation}
    \rho=\bigoplus_{F} p_F\; \rho^{(F)},
\end{equation}
where $p_F$ is the effective fraction of atoms having the total angular momentum $F$. Moreover, one may then focus on the dynamics of only one particular $\rho^{(F)}$ for the $F$-level actually contributing to the light-atom interaction, with $\rho^{(F)}$ being then decomposable just as in \eqnref{eq:rho_m}. In such a case, as done in what follows, the $(F)$-superscript can be dropped for simplicity.

As the case of a \emph{linearly polarised} pump is of our interest, see \figref{fig:geometry}, from symmetry arguments it follows that, independently of the spin-number $F$, only even-$\kappa$ coefficients are modified when interacting with light~\cite{Bevilacqua2014}. Moreover, as we show later, our model predicts no significant coupling between components of different $\kappa$. On the other hand, it is only the \emph{orientation} ($\kappa=1$) and \emph{alignment} ($\kappa=2$) components that can be probed by resorting to electric dipole light-atom interactions~\cite{Happer1972,Bevilacqua2014}. Hence, even though the pump in the experiment is relatively strong, so that the atomic state in \eqnref{eq:rho_m} reads:
\begin{equation}
    \rho=\tfrac{1}{2F+1}\openone_{2F+1}+ \sum_{q=-2}^{2}m_{2 q}\Ttens^{(2)}_q + \sum_{q=-4}^{4}m_{4 q}\Ttens^{(4)}_q +\dots ,
    \label{eq:rho_relevant}
\end{equation} 
any detection signal obtained by probing the atoms with light is determined by the vector containing the alignment coefficients:
\begin{equation}
    \mvec{m}=\left(
    \begin{array}{ccccc}
    m_{2,-2} & m_{2,-1} & m_{2,0} & m_{2,1} &  m_{2,2}
    \end{array}
    \right)^\TT,
    \label{eq:m_vec}
\end{equation}
whose dynamics must therefore only be tracked.

Furthermore, if the quantisation axis $z$ is chosen along the light polarisation of the pump, $\theta=0$ in \figref{fig:geometry}(a), by symmetry the atomic steady state must be invariant to any rotations around $z$. Hence, only the (real) coefficient with $q=0$ can acquire some value $m_{2,0}\to m^{\mrm{ini}}_{20}$, whose maximum (or negative minimum) is theoretically constrained by the positivity of $\rho\ge0$, but practically by the efficiency of optical pumping being counteracted by relaxation. Considering the light to be linearly polarised at an arbitrary angle to the $xz$-plane, see \figref{fig:geometry}(c), the steady state can be found by just adequately rotating the above solution for $\theta\!=\!0$ around the light-propagation direction $x$. In particular, the $\mvec{m}$-vector \eref{eq:m_vec} of the \emph{steady state} (ss) generated by linearly polarised pump at an angle $\theta$ with respect to the quantisation axis reads, see \appref{app:spherical_tensors}:
\begin{equation}
    \mvec{m}^\ss= m^{\mrm{ini}}_{20} 
    \begin{pmatrix}
        -\frac{\sqrt{6}}{4} \mathsf{s}_\theta^2 & \ii\frac{\sqrt{6}}{4} \mathsf{s}_{2\theta} & 1-\frac{3}{2}\mathsf{s}_\theta^2 & \ii\frac{\sqrt{6}}{4}\mathsf{s}_{2\theta} & -\frac{\sqrt{6}}{4}\mathsf{s}_\theta^2
    \end{pmatrix}^\TT,
    \label{eq:m_ss_exact}
\end{equation}
where $\mathsf{s}_\theta\coloneqq\sin\theta$. However, as the strong static field $\vec{B}_0$ leads to (Larmor) precession of the atomic state around $z$, see \figref{fig:geometry}, that is much faster than the timescale of reaching the steady state,  i.e.~with the Larmor (angular) frequency $\Omega_L\coloneqq\gmr|\vec{B}_0|$ much greater than the overall relaxation rate, all the multipoles $m_{2,q\neq 0}$ quickly average to zero, so that according to the secular approximation~\cite{Happer1972} the steady-state vector \eref{eq:m_ss_exact} simplifies to
\begin{equation}
    \mvec{m}^\ss \approx m^{\mrm{ini}}_{20} 
    \begin{pmatrix}
        0 & 0 & 1\!-\!\frac{3}{2}\sin^2\theta & 0 & 0
    \end{pmatrix}^\TT,
    \label{eq:m_ss}
\end{equation}
which we assume to be valid throughout this work. The Larmor frequency above is defined using the gyromagnetic ratio $\gmr \coloneqq g_F\mu_B/\hbar$ (with units $[\mathrm{rad}\,\mathrm{s}^{-1}\mathrm{T}^{-1}]$), where $g_F$ is the Land{\'e} g-factor for an atom of total spin $F$ and $\mu_B$ is the Bohr magneton.

In order to visualise the symmetries and geometric properties of the steady state \eref{eq:m_ss}, we resort to plotting the \emph{angular-momentum probability surfaces} it yields for $\theta=0$, $\theta=\tfrac{\pi}{2}$ and $0<\theta<\tfrac{\pi}{2}$ in \figref{fig:geometry}(a), (b) and (c), respectively. In particular, in each case we present a spherical plot of the overlap of the steady state with the state of a maximum angular momentum $\ket{F,F}_{\vec{n}}$ defined with respect to a given direction $\vec{n}$, which then determines the quantisation axis, i.e.:
\begin{equation}
    r(\vec{n})=\,_{\vec{n}}\!\bra{F,F}\rho_\ss \ket{F,F}_{\vec{n}},
\end{equation}
where $\rho_\ss$ is the steady-state density matrix of the form \eref{eq:rho_relevant} with the alignment coefficients given by \eqnref{eq:m_ss}.

In our experiment, as shown in \figref{fig:geometry}(c), the angular-momentum probability surface of the steady state is constantly perturbed out of equilibrium by the $\BrfVec$-field  inducing white noise in the $x$-direction, so that coefficients with $q\neq0$ in \eqnref{eq:m_ss} are no longer zero. As a result, the effective atomic state contains components not only from $\Ttens^{(2)}_0$, but also from $\Ttens^{(2)}_{\pm 1}$ and $\Ttens^{(2)}_{\pm 2}$ tensor operators, see \eqnref{eq:rho_relevant}, which can be separately visualised by the surfaces depicted in \figref{fig:geometry}(d). Crucially, as the latter two return to their original state under Larmor precession after being rotated by $2\pi$ and $\pi$, respectively, the measured noisy $\theta'$-signal should contain distinct frequency-components at $\Omega_L$ and $2\Omega_L$. This should be visible when analysing the \emph{power spectral density} (PSD) of the detected signal, as schematically presented in \figref{fig:geometry}(d).

\subsection{Magnetometer dynamics and measurement} \label{sec:dynamics}
In order to be able to predict the PSD in our experiment, we must move away from just the steady-state description. In particular, we must be able to model the stochastic dynamics of the atoms, so that the auto-correlation function of the detected signal can be computed, whose Fourier transform specifies the PSD.

\subsubsection{Atomic stochastic dynamics}

\paragraph{Dissipative non-unitary evolution.}
A dissipative evolution of the atomic state \eref{eq:rho_FM} is generally described by the Gorini–Kossakowski–Sudarshan–Lindblad equation~\cite{Breuer2002}:
\begin{equation}
    \frac{\dd\rho}{\dd t}=-\frac{\ii}{\hbar}[\hat H,\rho]+\Phi[\rho], 
    \label{eq:rho_evolution_det}
\end{equation}
where $\hat H$ is the system Hamiltonian, while the map
\begin{equation}
    \Phi[\rho]=\sum_i \GG_i (\hat L_i \rho \hat L_i^\dag-\frac 1 2 \{ \hat L_i^\dag \hat L_i, \rho \} )
\end{equation}
is responsible for the decoherence, with $\hat L_i$ being the quantum jump (Lindblad) operators and $\GG_i\ge0$ the corresponding dissipation rates, which must be non-negative for a Markovian evolution~\cite{Breuer2002}.

In the absence of the $\BrfVec$-field, the Hamiltonian incorporates only the interaction of the atom with the static field, i.e.~$\hat H = \gmr \vec{B}_0 \cdot \hat{\vec{F}} = \Omega_L \hat{F}_z$. However, as the static field introduces anisotropicity in the system, we split the decoherence map as follows:
\begin{equation}
   \Phi[\rho] \coloneqq \sum_{\alpha=x,y,z}\Phi_\alpha[\rho]+\Phi_{\mrm{iso}}[\rho],
   \label{eq:Phi}
\end{equation}
where
\begin{align}
    \Phi_\alpha[\rho] &\coloneqq \frac{\GG_\alpha}{\hbar^2} (\hat F_\alpha \rho \hat F_\alpha-\frac 1 2 \{ \hat F_\alpha^2, \rho \} )
\end{align}
can be interpreted as arising from magnetic-field fluctuation in each $\alpha=x,y,z$ direction, whereas:
\begin{align}
    \Phi_{\mrm{iso}}[\rho]& \coloneqq \rmdn{\GG}{iso}\!\left(\sum_{M,M'}\!\!\hat{L}_{M,M'}\rho\hat{L}_{M,M'}^\dagger \!-\!\frac{1}{2}\{\hat{L}_{M,M'}^\dagger\hat{L}_{M,M'},\rho\} \right)
\end{align}
with $\hat{L}_{M,M'} \coloneqq \frac{1}{\sqrt{2F+1}}\ketbra{F,M}{F,M'}$, represents isotropic dissipation that can be equivalently written as
\begin{equation}
    \Phi_{\mrm{iso}}[\rho]=\Lambda-\frac 1 2\{\hat\GG,\rho \},
    \label{eq:Phi_iso}
\end{equation}
where $\hat \GG \coloneqq \rmdn{\GG}{iso} \id_{2F+1}$ and $\Lambda \coloneqq\frac{\rmdn{\GG}{iso}}{2F+1} \, \id_{2F+1}$ 
are typically referred to as the re-population and depolarising terms, respectively~\cite{auzinsh_budker_rochester_2014}. Although we assume in our model the rates $\GG_\alpha$ and $\GG_\mrm{iso}$ to be phenomenological and account for various dissipation mechanisms, \eqnref{eq:Phi_iso} can be naturally interpreted as the loss of polarised atoms that then reappear in the beam in a completely unpolarised state.

\paragraph{Impact of the stochastic $\BrfVec$-field.}
In our experiment, see \figref{fig:geometry}(c), stochastic $\BrfVec$-field is applied in the light-propagation direction $x$, which leads to another term in the Hamiltonian $\Hrf(t) = \gmr \BrfVec(t) \cdot \hat{\vec{F}}= \gmr \Brf(t) \hat{F}_x$ with
\begin{equation}
     \Brf(t)= \frac{\Omegarf}{\gmr}\,\xi(t) \;\approx\; \frac{\Omegarf}{\gmr} \sqrt{\frac{1}{2\pi \Delta f}}\frac{\dd W_t}{\dd t},
     \label{eq:Bnoise}
\end{equation}
where $\Omegarf$ is an effective magnetic noise amplitude given in the units of Larmor frequency, while $\xi(t)$ represents the stochastic process for the noise we generate, see \appref{app:peak_amp} for its further characteristics, which effectively exhibits a constant power spectrum in a frequency range $f\in[\delta\!f,f_\trm{cutoff}]$ (in Hz) with $\delta\!f$ set close to zero to eliminate spurious low-frequency contributions, and some large cut-off $f_\trm{cutoff}$ imposed to prevent the effect of aliasing. Now, as we will deal with processes occurring at (angular) frequencies $\omega\approx\Omega_L$, such noise in \eqnref{eq:Bnoise} can be effectively described as white with the correct rescaling factor $\Delta f\coloneqq f_\trm{cutoff}-\delta\!f\approx f_\trm{cutoff}$---the white noise can be interpreted as the time-derivative, $\frac{\dd W}{\dd t}$, of the Wiener process, $W_t$~\cite{gardiner1985handbook}. As a result, we can write the (stochastic) time-increment induced by the noise involving the Hamiltonian as
\begin{equation}
   \Hrf(t) \dd t = \sqrt{\omegarf} \, \hat{F}_x \, \dd W_t,
\end{equation}
where we define $\omegarf \coloneqq \Omegarf^2/(2\pi f_\trm{cutoff})$ as the effective \emph{noise spectral density}, while $\dd W_t\sim\mathcal{N}(0,\dd t)$ is then normally distributed with variance $\dd t$, i.e.~the Wiener increment~\cite{gardiner1985handbook}.

In order to correctly include the white noise in the deterministic dynamics \eref{eq:rho_evolution_det}, we must explicitly compute the time-increment of the density matrix, $\dd \rho= \rho(t+\dd t)-\rho(t)$, that is now stochastic. By adding the noise contribution to \eqnref{eq:rho_evolution_det}, we define the stochastic map
\begin{equation}
    \mathcal{K}(t)[\rho] \coloneqq -\frac{\ii}{\hbar}[\Hrf(t)+\hat H,\rho]+\Phi[\rho],
\end{equation}
which allows us to write for small $\dd t$:
\begin{align}
    \rho(t+\dd t) & = \ee^{\mathcal{K}(t) \dd t}\left[\rho(t)\right]=\sum_{n=0}^\infty \frac{(\mathcal{K}(t) \dd t)^n[\rho(t)]}{n!} \nonumber \\ 
     & = \rho(t)+\mathcal{K}[\rho(t)]\dd t+\frac{1}{2}(\mathcal{K}\dd t)^2[\rho(t)] + O(\dd t^{5/2})\nonumber\\
     & = \rho(t)-\frac{\ii}{\hbar}[\hat H,\rho(t)]\dd t+\Phi[\rho(t)]\dd t \nonumber \\
     & -\frac{\ii\sqrt{\omegarf}}{\hbar}[\hat{F}_x,\rho(t)]\dd W_t-\frac{\omegarf}{2 \hbar^2}[\hat{F}_x,[\hat{F}_x,\rho(t)]]\dd t \nonumber \\
     & + O(\dd t^{3/2}),
\end{align}
where according to the It\^{o} calculus implying $\dd W_t^2=\dd t$ we obtain a dissipative term at the second ($n=2$) order, while all the other terms can be ignored with $\dd W_t \dd t = O(\dd t^{3/2})$ within the big-$O$ notation~\cite{gardiner1985handbook}.

As a consequence, we obtain the desired stochastic differential equation describing the atomic dynamics as
\begin{align}
    \dd\rho & = (-\frac{\ii}{\hbar}[\hat H,\rho]+\Phi_{\mrm{iso}}[\rho]) \dd t \nonumber \\
            & +\sum_{\alpha=y,z}\frac{\GG_\alpha}{\hbar^2}(\hat{F}_\alpha\rho\hat{F}_\alpha-\frac{1}{2}\{\hat{F}_\alpha^2,\rho\}) \dd t \nonumber \\
            & +\frac{\GG_x+\omegarf}{\hbar^2}(\hat{F}_x\rho\hat{F}_x-\frac{1}{2}\{\hat{F}_x^2,\rho\}) \dd t \nonumber \\
            & - \frac{\ii\sqrt{\omegarf}}{\hbar}[\hat{F}_x,\rho]\dd W_t,
    \label{eq:rho_evolution_stoch}
\end{align}
which, apart from the expected term generating random rotations around the $\BrfVec$-field direction, accounts for the fact that (by the fluctuation-dissipation theorem) the white noise must also increase the dissipation rate in the $x$-direction from $\GG_x$ to $\GG_x + \omegarf$.

\subsubsection{Spherical-tensor representation}
We have argued that when considering the relevant $F$-subspace of the atomic steady state described in \eqnref{eq:rho_m}, the nature and geometry of light-atom interactions allows us to reduce its form, so that it contains only the alignment component ($\kappa=2$) in \eqnref{eq:rho_relevant}. Consistently, as shown in \appref{app:spherical_tensors}, the evolution determined by \eqnref{eq:rho_evolution_stoch} does not couple spherical-tensor components of different rank $\kappa$. Hence, while incorporating optical pumping into the dynamics \eref{eq:rho_evolution_stoch}, the evolution of the atomic state must be completely described by the $\mvec{m}$-vector \eqref{eq:m_vec} of, now time-dependent, alignment coefficients, i.e.:
\begin{equation}
    \mvec{m}_t=\left(
    \begin{array}{ccccc}
    m_{2,-2}(t) & m_{2,-1}(t) & m_{2,0}(t) & m_{2,1}(t) &  m_{2,2}(t)
    \end{array}
    \right)^\TT.
    \label{eq:m_vect}
\end{equation}
This evolves under the dynamics \eref{eq:rho_evolution_stoch} according to the following stochastic differential equation, see \appref{app:spherical_tensors}:
\begin{align}
\dd\mvec{m}_t  
    & = \left(\mmat{A}_0 + \mmat{A}_{\Phi}+\frac{\rmdn{\mmat A}{\noise}^2}{2}\right) \mvec{m}_t\, \dd t + \rmdn{\mmat A}{\noise}\, \mvec{m}_t\, \dd W_t , 
    \label{eq:m_evolution1}
\end{align}
where $\mmat{A}_0 =- \ii \Omega_L \mmat{J}^{(2)}_z$ and $\rmdn{\mmat{A}}{\noise} =- \ii \sqrt{\omegarf} \mmat{J}^{(2)}_x$ are $5\times5$ matrices that should be associated with the free evolution and stochastic noise, respectively. These are defined with help of the representation of angular momentum operators, $\mmat{J}^{(\kappa)}_\alpha$ with $\alpha=x,y,z$, acting on the vector space of alignment---equivalent to ones acting on the state-space of a spin-2 particle, written in the $\{ \ket{2,-2}, \dots ,\ket {2,2} \}$ basis to agree with \eqnref{eq:m_vect}.

In a similar fashion, the matrix-representation of the dissipative map $\Phi$ in \eqnref{eq:Phi}, see \appref{app:spherical_tensors}, reads
\begin{align}
    \mmat A_\Phi 
    &=
    -\frac 1 2 \sum_\alpha \GG_\alpha (\mmat{J}^{(2)}_\alpha)^2-\rmdn{\GG}{iso}\openone_5 \nonumber \\
    &\equiv
    -\trm{diag}\{\GG_2,\GG_1,\GG_0,\GG_1,\GG_2\},
    \label{eq:A_Phi}
\end{align}
which we, however, force above to have a diagonal form postulated in \refcite{weis2006theory} with three dissipation rates:
\begin{equation}
	\vec{\GG} \equiv \begin{pmatrix}\GG_0\\\GG_1\\\GG_2\\\end{pmatrix}
	\coloneqq 
	\begin{pmatrix}6 \GG_\perp+\rmdn{\GG}{iso}\\\GG_\parallel+ 5 \GG_\perp+\rmdn{\GG}{iso}\\4\GG_\parallel+ 2 \GG_\perp+\rmdn{\GG}{iso}\\\end{pmatrix}.
	\label{eq:effective_rates}
\end{equation}
Formally, this corresponds to the assumption that in the absence of the induced magnetic-field noise one should differentiate only between dephasing rates along, $\GG_\parallel \coloneqq \GG_z$, and perpendicular to, $\GG_\perp \coloneqq \GG_x=\GG_y$, the static field, while keeping the isotropic depolarisation rate, $\rmdn{\GG}{iso}$, as an independent parameter. The effective rates $\vec{\GG}$, see also \appref{app:spherical_tensors}, are then defined by reparametrising the problem as in \eqnref{eq:effective_rates}.

Finally, in order to include the impact of optical pumping in \eqnref{eq:m_evolution1}, we enforce that, in the absence of the noisy magnetic field, the $\mvec{m}_t$-vector \eqref{eq:m_vect} must converge with time to the steady state described in \secref{sec:steady_state}. This way, we obtain the desired stochastic dynamical equation for the vector of alignment coefficients as
\begin{align}
\dd\mvec{m}_t  
    & = \left(\mmat{A}_0 + \frac{\rmdn{\mmat A}{\noise}^2}{2} \right) \mvec{m}_t\, \dd t + \mmat{A}_{\Phi}(\mvec{m}_t-\mvec{m}^\ss) \, \dd t \nonumber \\
    & + \rmdn{\mmat A}{\noise} \mvec{m}_t\, \dd W_t,
    \label{eq:dx_full}
\end{align}
where the steady state $\mvec{m}^\ss$ is given by \eqnref{eq:m_ss}, being already averaged over the fast (Larmor) precession around $z$ and, hence, satisfying $\mmat{A}_0\mvec{m}^\ss=\mvec{0}$.

\subsubsection{Detected signal}
We depict the phenomenon of polarisation rotation in \figref{fig:geometry}(c), i.e.~the effect that the angle $\theta$ of the linearly polarised incoming beam is changed to $\theta'$ upon leaving the atomic cell. Treating the atomic ensemble as an optically thin sample, the change of the angle, $\Delta\theta\coloneqq \theta'-\theta$, obeys then~\cite{auzinsh_budker_rochester_2014, meraki2023zero}:
\begin{equation}
    \Delta\theta\propto\ii (\tilde m_{2,1}+\tilde m_{2,-1}),
    \label{eq:rot_angle}
\end{equation}
with the proportionality constant depending on the optical depth, interaction strength, light power etc. 

In the above, $\tilde m_{\kappa,q}$ are the alignment coefficients defined with the quantisation axis along the direction of incoming light polarisation, i.e.~tilted away by $\theta$ from $z$ in the $yz$-plane, see \figref{fig:geometry}(c). Hence, \eqnref{eq:rot_angle} can be re-expressed with help of the $\mvec{m}_t$-vector \eref{eq:m_vect} (defined with $z$ being the quantisation axis) by simply rotating $\mvec{m}_t$ by an angle $\theta$ around the light-propagation direction $x$, i.e.:
\begin{equation} 
		\label{eq:rotation_signal}
    \Delta\theta \propto \mvec{h}^\TT\,\mmat{D}^{(2)}_\theta \mvec{m}_t,
\end{equation}
where $\mvec{h}=(0,\ii,0,\ii,0)^\TT$ and $\mmat{D}^{(2)}_\theta=D^{(2)}_{m,m'}(-\pi/2,\theta,\pi/2)$ is the appropriate Wigner D-matrix, see \appref{app:spherical_tensors}. 

As a result, based on \eqnref{eq:rotation_signal}, we may write the detected signal of an alignment-based magnetometer as:
\begin{align}
    S(t)
    &=g_D\,\mvec{h}^\TT\,\mmat{D}^{(2)}_\theta\mvec{m}_t +\zeta(t) \nonumber \\
    &=g_D\,\tfrac{1}{2}[\sqrt 6 m_{2,0}(t)+m_{2,2}(t)+m_{2,-2}(t)]\sin(2\theta) \nonumber \\
    &+g_D\,\ii [m_{2,1}(t)+m_{2,-1}(t)]\cos(2\theta) +\zeta(t) ,  
    \label{eq:det_signal}
\end{align}
where $g_D$ is the effective proportionality constant, whereas $\zeta(t)$ denotes the detection noise, which is completely uncorrelated from the magnetic-field noise affecting the atom dynamics in \eqnref{eq:dx_full}. Moreover, as we assume here the impact of the measurement backaction exerted on the atomic state by the light to be ignorable~\cite{Deutsch2010,Troullinou2021}, $\zeta(t)$ shall not exhibit any correlations with any noise exhibited by the atoms.

It becomes clear from the expression \eref{eq:det_signal} that, as the signal depends on all the alignment coefficients with $q=0,\pm1,\pm2$, it must contain components that oscillate at frequencies $0$, $\Omega_L$ and $2\Omega_L$, respectively. In other words, the signal contains information about different spherical-tensor components of the atomic state, in particular the ones illustrated in \figref{fig:geometry}(d), each of which should yield a peak in the PSD at the corresponding frequency.

\begin{figure}[t!]
    \centering
    \includegraphics[width=\linewidth]{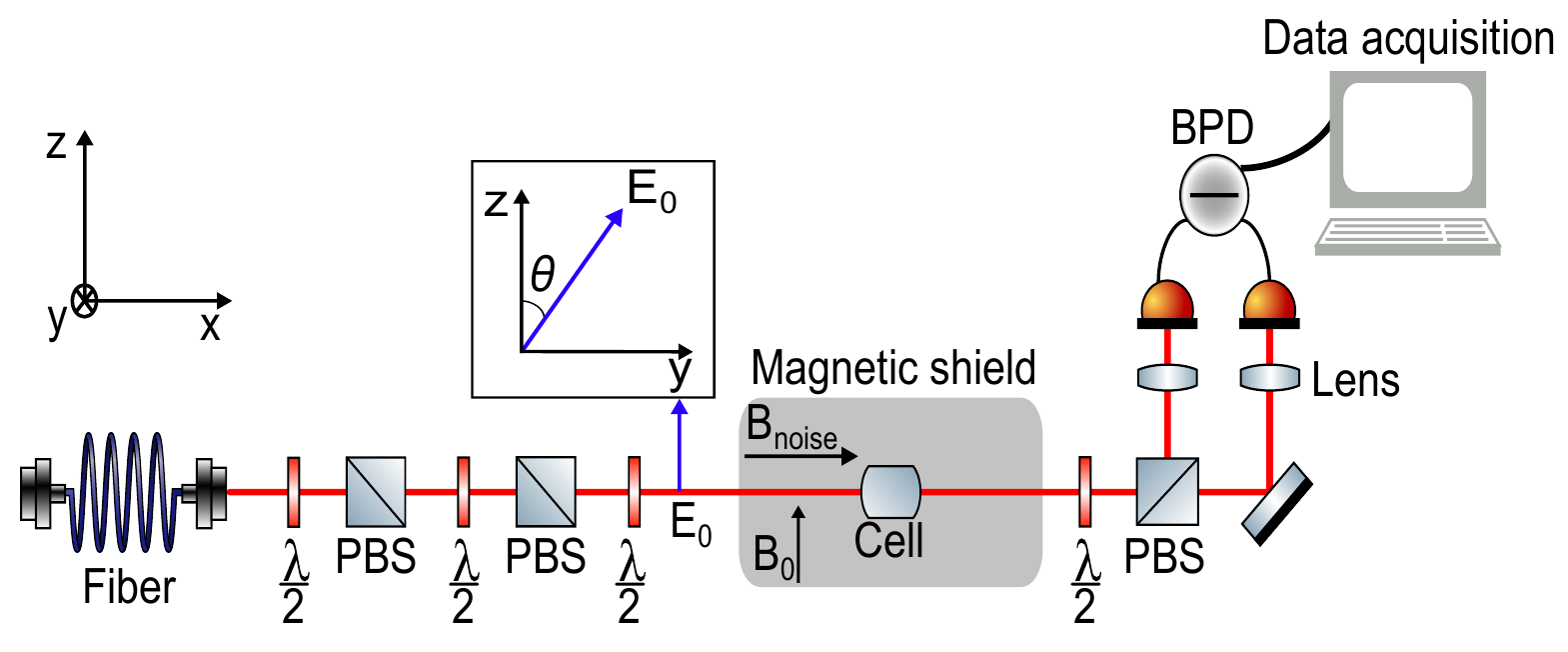}
    \caption{\textbf{Alignment-based magnetometer:~experimental setup}. Linearly polarised light is passed through a polarisation maintaining fibre. Two sets of half-wave plates ($\lambda /2$) and polarising beam splitters (PBS) are used to reduce intensity fluctuations and to change the power of the laser beam, respectively. The laser light propagates along the $x$-direction, while the angle $\theta$ of its linear polarisation is then adjusted by rotating yet another  half-wave plate, before it enters the magnetic shield (marked in grey). Within the shield, the light simultaneously pumps and probes the atomic ensemble of caesium atoms contained within a paraffin-coated cell (Cell). The atoms are subject to (strong) static and (relatively weak) noisy magnetic fields:~$\vec{B}_0$ applied in the $z$-, and $\BrfVec$ applied in the $x$-direction, respectively. Upon interacting with the atoms, the polarisation of light undergoes rotation, which is then measured by passing the output beam through a PBS and performing balanced photodetection (BPD).
  	}
    \label{fig:set_up}
\end{figure}

\subsection{Experimental setup} 
\label{sec:setup}
Figure~\ref{fig:set_up} shows a schematic of the experimental setup we employ. Linearly polarised light, with a wavelength of 895~nm, is passed through a polarization-maintaining optical fiber. The light is resonant with the $F=4\rightarrow F'=3$ D1 transition of caesium. The light has an electric field vector $\vec{E}_0$ and passes through a cubic paraffin-coated cell containing caesium atoms, which is placed inside a magnetic shield (Twinleaf MS-1). The cell is $(5 \text{mm})^3$ and is kept at room temperature ($\sim 18^{\circ}$). The laser beam diameter is $\sim2$~mm.

A static magnetic field $\vec{B}_0=B_0 \, \vec{e}_z$ is applied using the magnetic shield coils. A half-wave plate is placed before the magnetic shield to change the angle of the electric-field vector of the light, $\vec{E}_0$, with respect to the direction of the static magnetic field. Consistently with \figref{fig:geometry}(c), we denote in \figref{fig:set_up} by $\theta$ the angle between the linear polarisation and the $z$-direction in the $yz$-plane, so that, e.g., $\vec{E}_0 = E_0  \, \vec{e}_z$ for $\theta=0$ and $\vec{E}_0 = E_0  \, \vec{e}_y$ for $\theta=\pi/2$. In parallel, a noisy magnetic field $\BrfVec(t)=B_{\mathrm{\noise}}(t)\,\vec{e}_x$ is applied to the system using a home-made square Helmholtz coil. The current driving the coil corresponds to the white-noise signal outputted by a function generator, which is, however, first transformed thorough a $f_\trm{cutoff}=1$~MHz low-pass filter to prevent the effect of aliasing, as well as a $\delta\!f=1$~kHz high-pass filter to eliminate any spurious contributions at very low frequencies (see \appref{app:Calibration} for more details).

When measuring the power spectra of our magnetometer as a function of the strength of the noisy magnetic field, as described below, we set the static magnetic field to $B_0= 2.7~\mu$T, what corresponds to inducing the Larmor precession of the atomic spin at a frequency $f_L\approx 9.45$~kHz. In such a case, each dataset is acquired for a given fixed white-noise amplitude in the range of $70.91~\mathrm{nT_{rms}}$, for the first set, to $35.45~\mathrm{\mu T_{rms}}$, for the final dataset. When investigating the dependence on the polarisation angle $\theta$ instead, we use the static field $B_0= 2.74~\mu$T (i.e.~$f_L\approx 9.6$~kHz), while the white-noise amplitude is fixed to $1.4~\mathrm{\mu T_{rms}}$.

As shown in \figref{fig:set_up}, after the magnetic shield the polarisation rotation of the transmitted light is measured using a half-wave plate ($\lambda/2$) and a polarising beam splitter (PBS). The linear polarisation of the set up was checked before and after the vapor cell, by measuring the ellipticity of the beam which was found to be of negligible order. The half-wave plate is rotated to match exactly the polarisation angle $\theta'$ of the transmitted light, see \figref{fig:geometry}(c), so that it is the deviations from $\theta'$ that are then effectively measured via balanced photodection (BPD)%
\footnote{Even if $\theta'$ is not exactly matched, the resulting DC-component of the detected signal \eref{eq:det_signal} yields a spike at zero frequency in the PSD, whose presence may be safely ignored within our analysis.}.
In particular, the difference in the intensity of the outgoing beams from the PBS is tracked using a Thorlabs balanced photodetector (PDB210A/M). The output photocurrent is recorded in real time using a data acquisition card. Further technical details about the performance of the magnetometer used can be found in~\citeref{rushton2023alignment}. 

Importantly, as our experiment matches the spatial configuration of an alignment-based magnetometer described in previous sections, we can interpret the measured photocurrent of the BPD exactly as:
\begin{equation}
    \delta S(t)\coloneqq S(t)-\bar S, \label{eq:BPD_photocurrent}
\end{equation}
where by $\bar S \coloneqq \mean{S(t)}_{\mrm{ss}}$ we denote the time-independent mean DC component of the measured signal, which is determined by the mean of the steady-state solution of \eqnref{eq:dx_full}. As a result, \eqnref{eq:BPD_photocurrent} describes deviations from the mean value of the detected signal stated in \eqnref{eq:det_signal}. The effective proportionality constant $g_D$ in \eqnref{eq:det_signal}, which relates the instantaneous atomic state to the photocurrent signal, is then dictated by a number of experimental conditions including:~the light power (1~$\mu W$), pumping efficiency, size of the vapour cell ((5~mm)$^3$), as well as the temperature (room temperature, $\sim 18^{\circ}$). On the other hand, the detection noise, $\zeta(t)$ in \eqnref{eq:det_signal}, should be attributed to photon shot-noise and electronic jitter arising solely due to the photo-detection process that effectively leads to a background noise---within the measured PSD a DC offset is observed independently whether the light beam interacts with the atoms or not. This results in a noise floor that depends on the frequency, partially due to the 1/f noise. This will be taken into account when interpreting the data in \secref{sec:results} below.

\section{Spin noise spectroscopy} 
\label{sec:SNS}
Denoting the Fourier transform of any signal, here the measured current of the balanced photodetector $\delta S(t)$ defined in \eqnref{eq:BPD_photocurrent}, over a finite-time interval $[0,T]$ as:
\begin{equation}
    \delta S(\ww)=\frac 1 {\sqrt{T}}\int_0^{T} \dd t \ee^{-\ii \ww t} \delta S(t),
    \label{eq:signal}
\end{equation}
its \emph{power spectral density} (PSD) is defined as~\cite{sinitsyn2016theory}:
\begin{equation}
    \mrm{PSD}(\ww) \coloneqq \mean{| \delta S(\ww) |^2 },
    \label{eq:PSD}
\end{equation}
where by $\mean{\dots}$ we denote throughout the article an average over stochastic trajectories. 

Importantly, provided that $\delta S(t)$ is stationary and ergodic, which can be assured by letting $T\gg \tau_\mrm{coh}$ in \eqnref{eq:signal} where $\tau_\mrm{coh}$ is some effective coherence time of the noisy system under study, we may rewrite the power spectrum according to the Wiener-Khinchin theorem as~\cite{sinitsyn2016theory}:
\begin{subequations}
\begin{align}
    \mrm{PSD}(\ww)\; 
    &\underset{\trm{ss}}{=}\;
    2\int_0^{\infty} \dd t \cos (\ww t) \mean{\delta S(t)\delta S(0)}
    \label{eq:PSD_delta_S} \\
    &\underset{\trm{ss}}{=}\; 
    2\int_0^{\infty} \dd t \cos (\ww t) \mean{S(t),S(0)} 
    \label{eq:PSD_WK_theorem}
\end{align}
\end{subequations}
where by ``ss'', as before, we emphasise the above to hold in the steady state. In particular, see \eqnref{eq:PSD_delta_S}, the PSD \eref{eq:PSD} can be expressed in terms of the \emph{auto-correlation function} of the (zero-mean) signal $\delta S(t)$; or equivalently, see \eqnref{eq:PSD_WK_theorem}, the \emph{auto-covariance function} of the actual detected signal, $S(t)$ specified in \eqnref{eq:det_signal}, i.e.~$\mean{S(t_1),S(t_2)} \coloneqq \mean{S(t_1)S(t_2)}-\mean{S(t_1)}\mean{S(t_2)}$~\cite{gardiner1985handbook}.

Substituting explicitly the form of the detected signal \eqref{eq:det_signal} into \eqnref{eq:PSD_WK_theorem}, we obtain the form of the PSD applicable to our problem as
\begin{equation}
    \mrm{PSD}(\ww)=g_D^2\,\mvec{h}^\TT\,\mmat{D}^{(2)}_\theta \mbf{\Xi}(\ww)\mmat{D}^{(2)\TT}_\theta \mvec{h} + \mean{|\zeta(\ww)|^2},
    \label{eq:PSD_solution}
\end{equation}
where the (5x5) matrix $\mbf{\Xi}(\ww)$ is defined as:
\begin{equation}
    \Xi_{pq}(\ww) \coloneqq 2\int_0^{\infty} \dd t \cos (\ww t) \langle m_{2,p}(t),m_{2,q}(0) \rangle,
\end{equation}
with $p,q=-2,..,2$ specifying the coefficients of the $\mvec{m}_t$-vector \eref{eq:m_vect} evaluated in the steady state. As the detection noise $\zeta(\ww)$ is uncorrelated from any other noise sources within our model, it leads to a noise floor in \eqnref{eq:PSD} as expected.

\subsection{Theoretical predictions} 
\label{sec:predictions}
The alignment dynamics derived in \eqnref{eq:dx_full} constitutes an example of a stochastic inhomogeneous evolution~\cite{gardiner1985handbook}, for which one can explicitly determine the form of the $\bm \Xi (\ww)$-matrix appearing in the PSD \eref{eq:PSD}, see \appref{app:spectrum_from_equation}, i.e.:
\begin{equation}
    \bm \Xi(\ww) = (\rmdn{\mmat A}{det} + \ii \ww)^{-1} \rmdn{\mmat A}{\noise}\, \bm \sigma\, (\rmdn{\mmat A}{\noise})^\TT (\rmdn{\mmat A}{det}^\TT - \ii \ww)^{-1},
\end{equation}
where by $\rmdn{\mmat A}{det}\coloneqq\mmat{A}_0+\mmat{A}_{\Phi}+\tfrac{1}{2}\rmdn{\mmat A}{\noise}^2$ we denote for short the overall matrix responsible in \eqnref{eq:m_evolution1} for the deterministic evolution. The (covariance) matrix $\bm \sigma$ above is then, see \appref{app:spectrum_from_equation}, the solution of the \emph{linear} equation:
\begin{align}
    &\rmdn{\mmat A}{det} \bm \sigma + \bm \sigma \rmdn{\mmat A}{det}^\TT + \rmdn{\mmat A}{\noise} \, \bm \sigma \, (\rmdn{\mmat A}{\noise})^\TT \nonumber \\
    &= \mmat{A}_{\Phi} \mvec{m}^\ss  (\mvec{m}^{\ss})^{\TT}{\mmat{A}_{\Phi}}^\TT (\rmdn{\mmat A}{det}^\TT)^{-1}\nonumber \\
    &\qquad+(\rmdn{\mmat A}{det})^{-1} \mmat{A}_{\Phi} \mvec{m}^\ss  (\mvec{m}^{\ss})^{\TT}{\mmat{A}_{\Phi}}^\TT,
\end{align}
which can always be solved fast numerically, given $\mmat{A}_0$, $\rmdn{\mmat A}{\noise}$, $\rmdn{\mmat A}{\Phi}$ and $\mvec{m}^\ss$.

However, independently of the particular form of $\bm{\sigma}$, one can show that the PSD \eref{eq:PSD_solution} for our problem, see \appref{app:spec_form}, must correspond to a sum of absorptive and dispersive Lorentzian functions:
\begin{align}
		\label{eq:PSD_predicted}
    \mrm{PSD}(\ww)=& \sum_{j=-2,..,2}\frac{p_{|j|}^a\gam_{|j|}^2}{(\ww-\omega_j)^2+\gam_{|j|}^2} \\
    &+ \sum_{j=\pm 1,\pm 2}\frac{\pm p_{|j|}^d\gam_{|j|}(\ww-\omega_j)}{(\ww-\omega_j)^2+\gam_{|j|}^2} + \mean{|\zeta (\ww)|^2}, \nonumber
\end{align}
whose \emph{central frequencies}, $\omega_j$, read
\begin{align}
    \ww_0&=0, \quad \ww_1=-\ww_{-1}=\sqrt{\Omega_L^2-\frac{9}{16}\omegarf^2}, \nonumber \\
    \ww_2&=-\ww_{-2}\approx 2\Omega_L-\frac 3 {16} \frac{\omegarf^2}{\Omega_L},
    \label{eq:omega_j}
\end{align}
and, as expected, up to negligible corrections $O(\omegarf^2/\Omega_L)$ correspond to multiples of the Larmor frequency:~$0$, $\Omega_L$ and $2\Omega_L$. Whereas, the \emph{line widths} (half-widths at half maxima), $\gamma_{|j|}$, take the form
\begin{align}
    \gam_0&\approx \GG_0+\frac 3 2 \omegarf, \quad \gam_1=\GG_1+\frac 5 4 \omegarf, \nonumber\\
    \gam_2&\approx\GG_2+\frac 1 2 \omegarf,
    \label{eq:gamma_j}
\end{align}
with $\Gamma_i$ defined as in \eqnref{eq:effective_rates}. All $\omega_j$ and $\gamma_{|j|}$ stated in \eqnsref{eq:omega_j}{eq:gamma_j}, respectively, can be determined analytically, however, we already simplified their forms above for  $\omega_{\pm2}$, $\gamma_0$ and $\gamma_2$, which are given by $\forall_j:\;\Omega_L\gg \Gamma_{|j|}$ and $\Omega_L\gg \omegarf$.

In particular, these assumptions are guaranteed in our experiment, in which the static field $B_0$ is always much stronger than the noisy field $B_\mrm{\noise}$ and yields the Larmor frequency $\Omega_L$ much greater than any of the dissipation rates forming $\bm {\Gamma}$ in \eqnref{eq:effective_rates}. Moreover, under these assumptions we can compute analytically also the \emph{peak heights}, $p_{|j|}^{a/d}$, which in the absorptive case then read
\begin{subequations}
\label{eqn:peaks_omega}
	\begin{equation}\label{eqn:peaks_0}
		p^a_0 = C\, \frac{27}{16}\;\omegarf^2\;\frac{\GG_0^2(2\Gamma_2+\omegarf)}{(2\GG_0+3 \omegarf)^2 G(\omegarf,\vec{\GG})} \; h(\theta),
	\end{equation}
  \begin{equation}\label{eqn:peaks_omega1}
  	p^a_1 = C\, \frac{3}{4}\;\omegarf\;\frac{\GG_0^2(2\Gamma_2+\omegarf)}{(4\GG_1+5 \omegarf)\,G(\omegarf,\vec{\GG})} \; g(\theta),
  \end{equation}
  \begin{equation}
  	p^a_2 = C\, \frac{3}{32}\;\omegarf^2\;\frac{\GG_0^2}{(2\GG_2+ \omegarf)\,G(\omegarf,\vec{\GG})} \; h(\theta),
  	\label{eqn:peaks_omega2}
  \end{equation}
\end{subequations}
with 
\begin{align}
    &G(\omegarf,\vec{\GG}) \coloneqq (2\GG_0+3 \omegarf) \{ 3\omegarf^2[\GG_0+2(\GG_1+\GG_2)] \nonumber \\
    &\;\;+2\omegarf(2\GG_0\GG_1+5\GG_0\GG_1+6\GG_1\GG_2)+8\GG_0\GG_1\GG_2 \},
    \label{eq:G_function}
\end{align}
and the proportionality constant $C \coloneqq g_D^2(m^{\mrm{ini}}_{20})^4$. 

The dispersive equivalents of expressions \eref{eqn:peaks_omega} can also be determined analytically (given $\forall_j:\;\Omega_L\gg \Gamma_{|j|}$ and $\Omega_L\gg \omegarf$) and can be found in \appref{app:peak_amp}. However, see \appref{app:disp_contr}, these are negligible upon substituting the parameters applicable to our experiment. Hence, we ignore their contribution to the PSD \eref{eq:PSD_predicted} from now on.

Although Eqs.~\eref{eqn:peaks_omega} allow us to predict the dependence of the peak heights for all values of noise intensity $\omegarf$, it directly follows that for 
low noise-strengths they obey $p_0^a~\propto~\omegarf^2$, $p_1^a \propto \omegarf$, and $p_2^a \propto \omegarf^2$. Furthermore, their angular dependence factorises and is given by
\begin{equation}\label{eq:h_theta}
    h (\theta)  \coloneqq [2\sin(2\theta)+3\sin(4\theta)]^2
\end{equation}
for  $p_0^a$ and $p_2^a$, whereas for $p_1^a$ it reads
\begin{equation}\label{eq:g_theta}
      g (\theta)  \coloneqq [3+2\cos(2\theta)+3\cos(4\theta)]^2.
\end{equation}
Consistently, these angular dependencies correspond to the squares of the expressions derived in \citeref{akbar2024optimized}, when the response to a radio-frequency magnetic field is considered instead of white noise.

\subsection{Measured noise spectra} 
\label{sec:results}
In order to validate the theoretical model outlined above, we vary the noisy magnetic field in the table-top alignment-based magnetometer described in \secref{sec:setup}. For the purpose of experiment, we define the noise spectral density in Hz, i.e.~$\frf$ such that $\omegarf = 2 \pi \frf$. As we apply a known voltage through the coil, it is convenient to further write $\frf = c\, \Vrf^2$, where $\Vrf$ has units of $\mathrm{mV_{rms}}$ and the proportionality constant $c$ can be explicitly determined for our setup, see \appref{app:Calibration}. 

\begin{figure}[t!]
    \centering
    \includegraphics[width=\columnwidth]{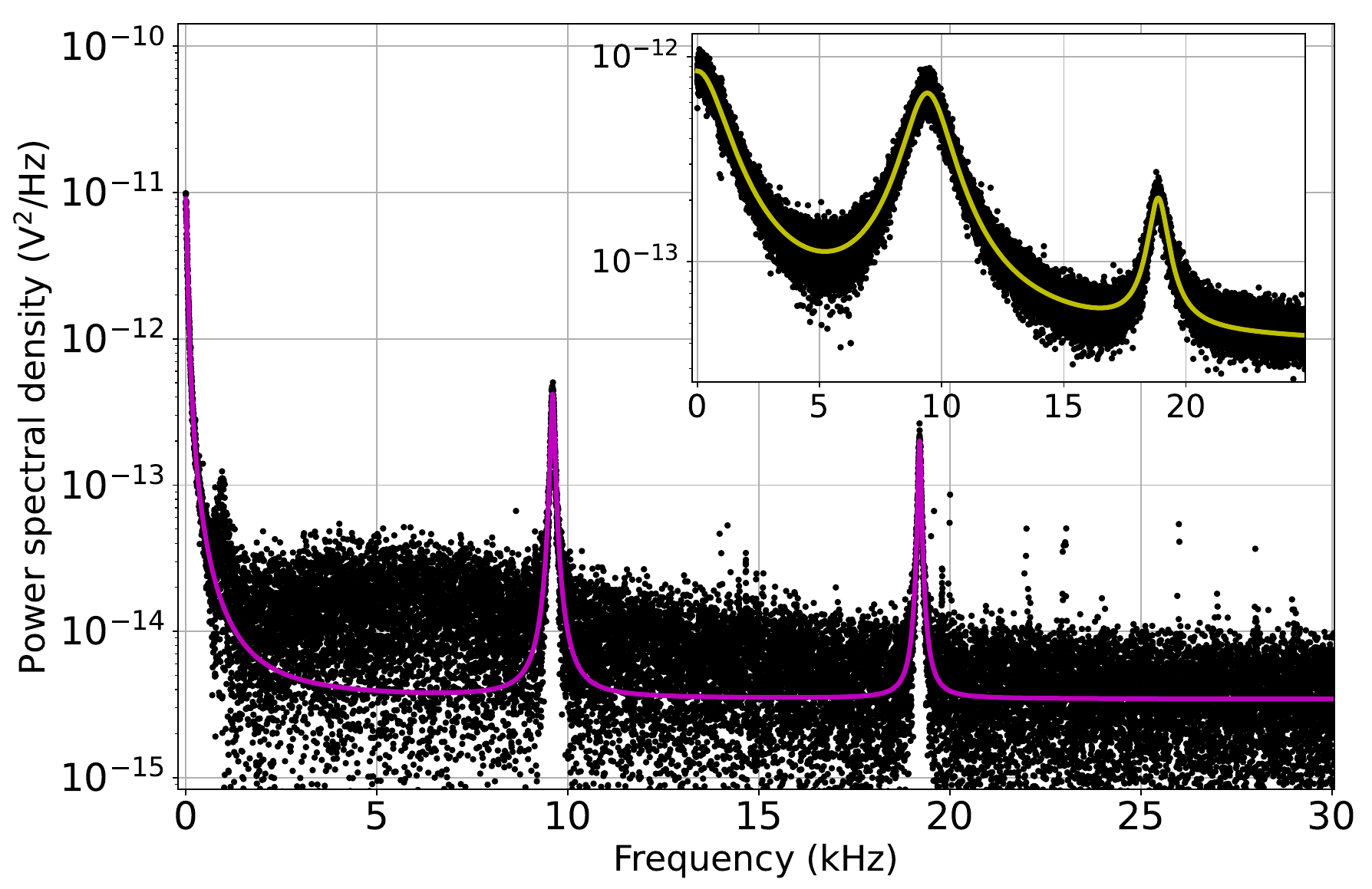}
    \caption{
    \textbf{Power spectral density} (PSD) measured in the experiment against the theoretical fits predicted by \eqnref{eq:PSD_predicted}. The signal was recorded over 100 time traces, each of 1 second duration. Within the main plot, relatively \emph{low-noise} spectral density of $\frf = 0.26$~Hz is chosen, corresponding to 140~mV$_\mathrm{rms}$ (the input polarisation angle is set to $\theta=40^\circ$). The so-obtained PSD has the experimental noise floor subtracted and is then fitted using a single function (magenta curve) containing three Lorentzian peaks with centres located close to the frequencies:~$f = 0$, $f_L$ and $2f_L$. Here, the three peaks are clearly distinguishable due to their small linewidths. In contrast, the inset shows the PSD for \emph{high-noise} spectral density $\frf = 120$~Hz, corresponding to 3~V$_\mathrm{rms}$ (with the input polarisation angle set now to $\theta=25^\circ$), that nonetheless is well described by the same fit-function (yellow curve) despite the Lorentzian peaks now overlapping significantly.
    }
    \label{fig:PSD_fits}
\end{figure}

In order to measure the PSD \eref{eq:PSD} of our device, we record for a given value of $\frf$ one hundred one-second datasets, for each of which the Fourier transform is then computed before averaging. The main plot of \figref{fig:PSD_fits} shows an exemplary PSD obtained for the input polarisation angle being set to $\theta=40^{\circ}$, and a relatively low noise spectral density being applied, $\frf = 0.26$~Hz (corresponding to $\Vrf = 140$~mV$_{\mathrm{rms}}$). It is clear that the PSD has peaks at approximately $0$, $f_L$ and $2f_L$ frequencies with $\Omega_L = 2 \pi f_L$, as anticipated by \eqnref{eq:omega_j}. 

Due to a clear separation of the peaks, the noise floor varies between them---it takes the value of approximately $6.5 \times 10^{-14}$~V$^2$/Hz for the peaks centred at $f_0 = 0$ and $f_1\approx f_L$, while it reads about $2.7 \times 10^{-14}$~V$^2$/Hz for the peak at $f_2\approx 2 f_L$. When fitting the data we subtract the noise floor (data not shown) of the experiment from each data set. This removes the detection noise (and $1/f$-noise) of different strengths across the spectrum. We then fit a single function containing three Lorentzian peaks to the whole spectrum, i.e.~the complete absorptive part of \eqnref{eq:PSD_predicted}. The corresponding fit parameters obtained for each of the three peaks are listed in \tabref{tab:individual}. In contrast, when dealing with \emph{high strengths of white noise}, the three peaks strongly overlap within the PSD, as shown explicitly for $\frf=120~\mathrm{Hz}$ within the inset of \figref{fig:PSD_fits}. As predicted by the theory, this is due to an apparent increase of the peak linewidths. 

\begin{figure*}[t!]  
		\includegraphics[width=0.95\linewidth]{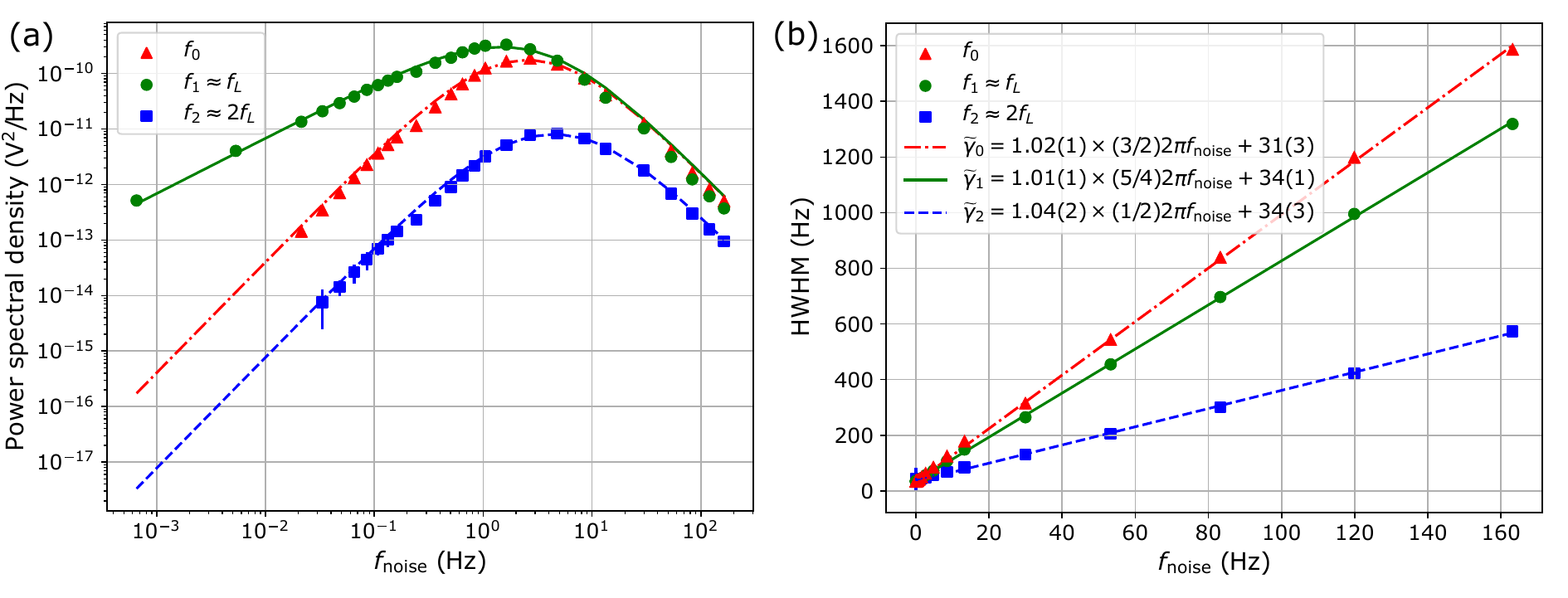}
    \caption{\textbf{Amplitudes (a) and linewidths (b) of the PSD peaks as a function of the white-noise strength}, which is varied from $6.53 \times 10^{-4}$~Hz to 163~Hz with the light polarisation angle fixed to $\theta = 25^{\circ}$. For every white-noise value $f_{\mathrm{noise}}$, amplitudes and linewidths are determined for each of the three relevant (absorptive) peaks in the PSD via the fitting procedure described in \figref{fig:PSD_fits}. The theoretical predictions are shown for each peak at $0$ (red, dot-dashed), $\approx f_L$ (green, solid) and $\approx 2f_L$ (blue, dashed) frequency. (a)~All three peak amplitudes $\widetilde{p}_j^a$ clearly follow the theoretical predictions of \eqnref{eqn:peaks_omega}, which we fit by tuning the overall proportionality constants (allowing different $C_j$ for each $p^a_j$ in \eqnref{eqn:peaks_omega}) and common values of the dissipation rates ($\Gamma_0$, $\Gamma_1$ and $\Gamma_2$ in \eqnref{eqn:peaks_omega}). (b)~The linewidths increase linearly with $\frf$ with the slope agreeing almost exactly for each peak with the proportionality constants predicted by \eqnref{eq:gamma_j}. Moreover, their offsets at $\frf=0$ provide us independently with the dissipation rates ($\Gamma_0$, $\Gamma_1$ and $\Gamma_2$ appearing also in \eqnref{eq:gamma_j}), which are consistent (up to $\approx\!10$Hz) with their equivalents predicted by amplitude-fitting in (a)---see the main text.
    }
    \label{fig:WN}
\end{figure*}

In what follows, we verify in more detail the expressions \eqref{eqn:peaks_omega} for the peak amplitudes by studying explicitly their dependence on the noise spectral density, $\frf$, and the light polarisation angle, $\theta$.

\begin{table}[t!]
\begin{tabular}{ | m{0.6cm}| m{1.5cm} | m{1.5cm} | m{2.3cm}|} 
  \hline
   $j$ & $f_j$ & $\widetilde{\gamma}_j$ & $\widetilde{p}_{j}^a$ \\ 
   & (Hz) & (Hz) & (V$^2$/Hz)\\ 
  \hline
   $0$ & $0$ & $34.9(2)$ & $9.03(3) \times 10^{-12}$  \\ 
  \hline
    $1$ & $9603(1)$ & $48.6(4)$ & $4.16(2) \times 10^{-13}$ \\ 
  \hline
  $2$ & $19208(1)$ & $37.8(6)$ & $1.97(2) \times 10^{-13}$\\ 
  \hline
\end{tabular}
    \caption{
    \textbf{Fit parameters in the low-noise regime.} A single function corresponding to the absorptive part of \eqnref{eq:PSD_predicted} is fitted to reconstruct the PSD presented in the main plot of \figref{fig:PSD_fits}. As it contains three Lorentzian profiles ($j=0,1,2$), such a procedure yields three distinct sets of parameters:~central frequencies ($f_j$), linewidths ($\widetilde{\gamma}_j$) and peak amplitudes ($\widetilde{p}_{j}^a$). The data was collected in the above measurement of the PSD (main plot of \figref{fig:PSD_fits}) for Larmor frequency $\sim\!9.6$~kHz, the light polarisation angle being set to $\theta=40^\circ$, and the strength of applied noise set to 140~mV$_\mathrm{rms}$ such that $\frf = 0.26$~Hz.
    } 
    \label{tab:individual}
\end{table}

\subsubsection{Varying the noise spectral density}
\label{sec:validation}
Figures \ref{fig:WN}(a) and \ref{fig:WN}(b) show the variation of the amplitude and linewidth, respectively, for each of the Lorentzian peaks fitted around $f = 0$, $f_L$ and $2f_L$, as a function of the noise spectral density $\frf$ (in Hz). The data was collected with a fixed, $\theta = 25^{\circ}$, polarisation angle of the input beam. As described above, the fitting procedure was done using a single fitting-function. The applied voltage to generate the noisy magnetic field was varied from 7~$\mrm{mV_{rms}}$ to 3500~$\mrm{mV_{rms}}$. To convert this to spectral density (in Hz) we determine the proportionality constant in $\frf = c \mrm{V^2_{\noise}}$ as $c = 1.33(3) \times 10^{-5}~\mathrm{Hz/mV^2_{rms}}$, which is possible via the calibration procedure described in \appref{app:Calibration}.

Crucially, the experimental results are consistent with the theory, in particular, the amplitudes of the three (absorptive) peaks follow the functional dependences predicted by \eqnref{eqn:peaks_omega}, with e.g.~quadratic and linear dependences:~$\widetilde{p}_0^a~\propto~\frf^2$, $\widetilde{p}_1^a \propto \frf$, and $\widetilde{p}_2^a \propto \frf^2$;~easily verifiable from \figref{fig:WN}(a) in the low white-noise regime. Performing the full fitting procedure, we allow the proportionality constants in \eqnref{eqn:peaks_omega} to differ ($C_j$ for each $p^a_j$), while determining common dissipation rates that are most consistent with the data ($\Gamma_0$, $\Gamma_1$ and $\Gamma_2$). We obtain $C_0 = 2.1(2) \times 10^{-6}~\mathrm{V}^2$, $C_1 = 3.4(6) \times 10^{-6}~\mathrm{V}^2$ and $C_2=1.0(1) \times 10^{-6}~\mathrm{V}^2$ for each peak, respectively, which consistently are of the same order, whereas the best-fitted dissipation rates read $\widetilde{\Gamma}_0=29(1)$~Hz, $\widetilde{\Gamma}_1=43(4)$~Hz and $\widetilde{\Gamma}_2=29(2)$~Hz. On the other hand, the corresponding linewidths depicted in \figref{fig:WN}(b) follow linear dependences in $\frf$ with the slope almost identical to the proportionality constants predicted by \eqnref{eq:gamma_j}. Moreover, their offsets at $\frf$ allows us to independently determine the dissipation rates \eref{eq:effective_rates} as $\widetilde{\Gamma}_0=31(3)$~Hz, $\widetilde{\Gamma}_1=34(1)$~Hz and $\widetilde{\Gamma}_2=34(3)$~Hz. These are in a good agreement (within $\approx\!10$Hz) with the ones found when fitting the three peak-amplitudes in \figref{fig:WN}(a).

\begin{figure}[b!]
    \includegraphics[width=\columnwidth]{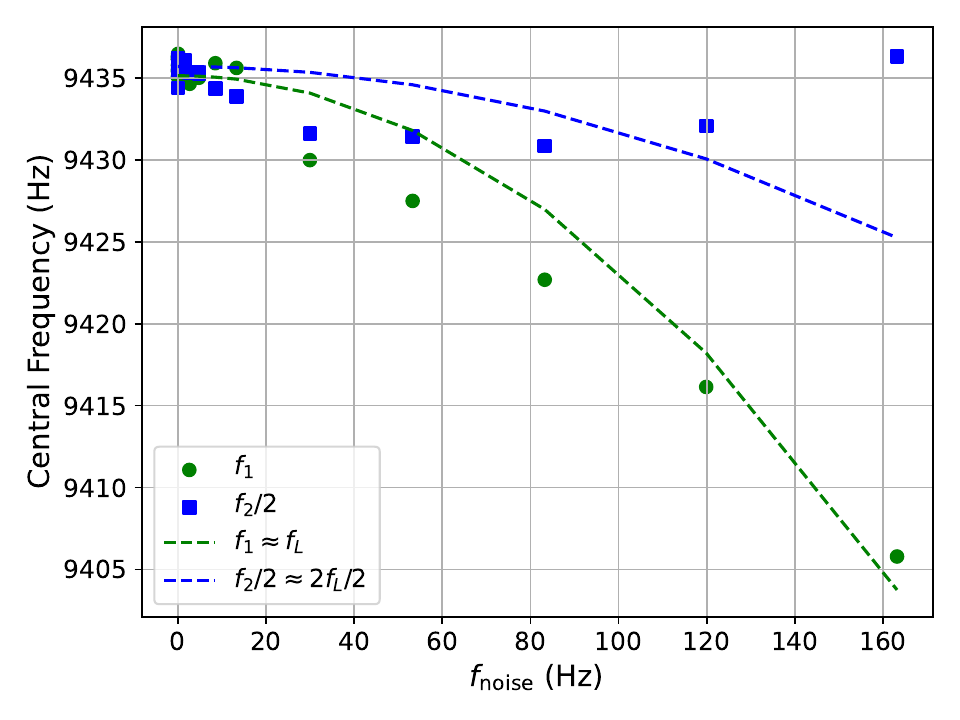}
    \caption{\textbf{Central frequency of the PSD peaks as a function of the noise spectral density ($\frf$)}, which is varied from $6.53 \times 10^{-4}$~Hz to 163~Hz with the light polarisation angle being fixed to $\theta = 25^{\circ}$. The theoretical predictions of the change in central frequency (\eqnref{eq:omega_j}) are shown over the full range of the noise spectral densities applied (dashed lines) with $f_L = 9435(1)$~Hz fitted to the experimental data for $f_1 \approx f_L$ (green dots) and $f_2 \approx 2f_L$ (blue squares). The overall error in the experimentally determined central frequencies is appr.~$\pm5$~Hz, arising both from experimental imperfections (e.g.~drifts, background-noise subtraction) and the fitting procedure.
    }    
    \label{fig:WN_freq}
\end{figure}

Last but not least, let us comment on the ability of our theoretical models to predict changes in the central frequencies of the peaks. \figref{fig:WN_freq} shows how the central frequency, for the peak around $f_L$ (green dots) and $2f_L$ (blue squares), vary as a function of the noise spectral density in the experiment. The theoretically predicted relation, \eqnref{eq:omega_j}, is then fit to the experimental data (dashed lines) with $f_L = \Omega_L/(2\pi) = 9435(1)$~Hz, being the only free parameter. It can be seen that the theory agrees reasonably well with the experimental data with the central frequency slightly decreasing in the high noise regime for the peak at $f_L$ and remaining almost the same for the second peak. Note that the error on the values for the central frequencies are within $\pm 5$~Hz. These errors come from experimental imperfections, however, such as the central frequency drifting in the time it takes for all of the data to be taken, and errors in fitting the data. It is noted here that the errors in fitting the central frequency at $2f_L$ will be larger than at $f_L$ due to the peaks being at least an order of magnitude smaller and hence closer to the noise floor of the experiment. Bearing in mind all the aforementioned imperfections, one may conclude that theory predicts well the dependence of the central frequencies on the noise density for both peaks.

\subsubsection{Varying the light polarisation angle}
\label{sec:validation2}
\begin{figure}[t!]
    \centering
    \includegraphics[width=\columnwidth]{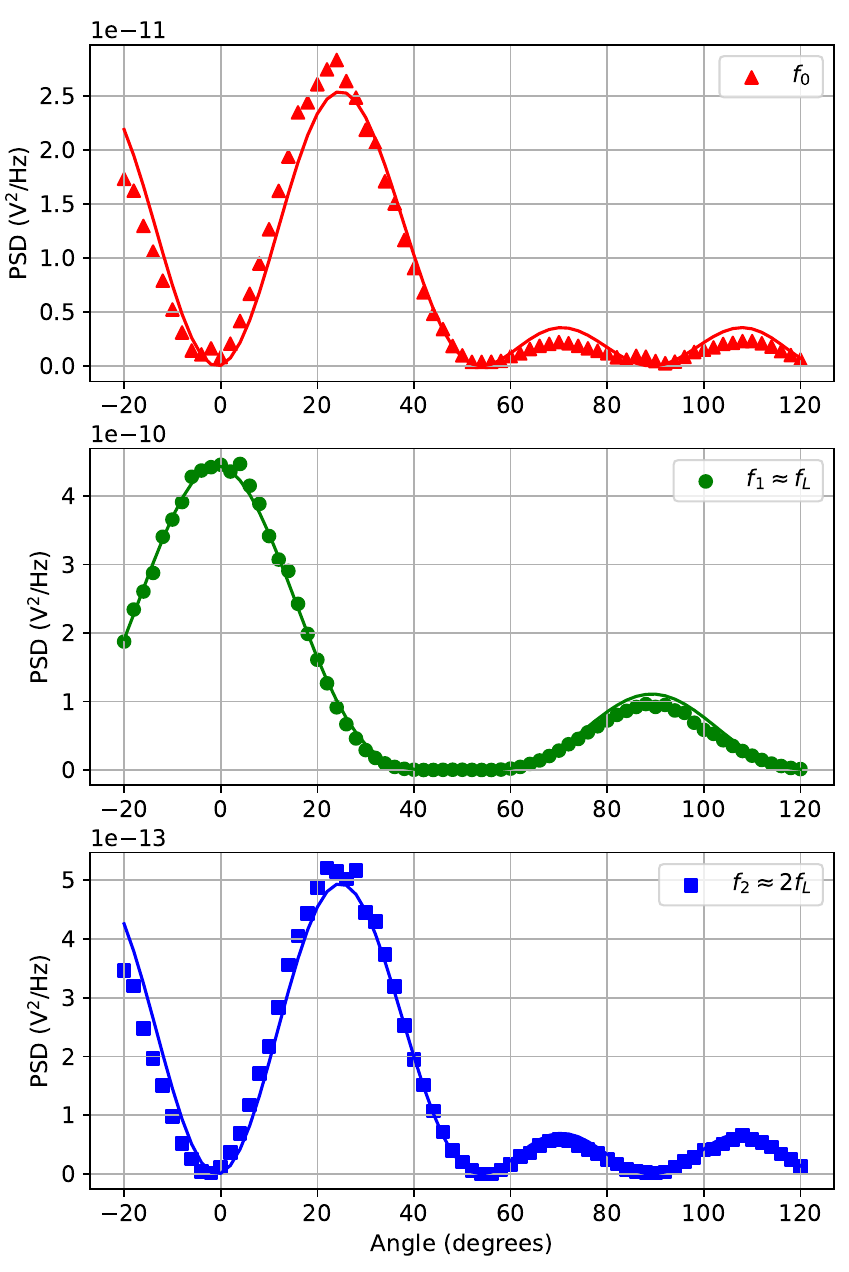}
    \caption{\textbf{Amplitudes of the PSD peaks as a function of the light polarisation angle}, which is varied from $\theta = -20^{\circ}$ to $\theta = 120^{\circ}$ with the white-noise strength set to $\Vrf=140~\mVrms$, corresponding to $\frf = 0.26~\mrm{Hz}$. The fits predicted theoretically by \eqnref{eqn:peaks_omega}, in particular, the angular dependences $h[\theta]$ and $g[\theta]$ stated in \eqnsref{eq:h_theta}{eq:g_theta}, are shown as solid curves for the peaks centred around frequencies $0$ (red), $f_L$ (green) and $2f_L$ (blue).}
    \label{fig:theta_ampl}
\end{figure}

Finally, as the angular dependence of the peak amplitudes \eref{eqn:peaks_omega} separates from all other parameters in the form of functions $h[\theta]$ and $g[\theta]$ stated in \eqnsref{eq:h_theta}{eq:g_theta}, we verify this explicitly in the low white-noise regime with $\Vrf = $140~$\mVrms$, corresponding to $\frf = 0.26~\mrm{Hz}$, by varying the polarisation angle of the incoming light from $\theta = -20^{\circ}$ to $\theta = 120^{\circ}$. It is noted that the Larmor frequency used for varying the polarisation angle was $\sim 9.6$~kHz. The results are presented in \figref{fig:theta_ampl} with measurements reproducing almost exactly the predicted angular behaviours. As the linewidths of the peaks do not vary significantly when varying $\theta$ (data not shown), we may use the values $\gamt_{j}$ determined for this low value of noise in \figref{fig:WN}(a) and compute separately the exact proportionality constants for each $\widetilde{p}^a_j$ such that the functions $h[\theta]$ and $g[\theta]$ are most accurately reproduced. In this way, while accounting also for a common angular offset $\theta \to \theta + \delta\theta$
with $\delta\theta = 0.77(7)^{\circ}$ in our setup, we obtain $C_0 = 3.9(6) \times 10^{-6}$~V, $C_1 = 2.1(1) \times 10^{-6}$~V and $C_2 =1.5(4)\times 10^{-6}$~V, which are consistently of similar magnitude and almost the same as the ones determined above when varying the white-noise strength.

\section{Conclusions} 
\label{sec:conclusions}
We prepare a detailed dynamical model allowing us to predict spin-noise spectra of an alignment-based magnetometer, which we then verify experimentally. Its applicability relies on the presence of the excess white-noise being applied along the propagation direction of the light beam that is used to both pump and probe the atomic ensemble, which is perpendicular to the direction of the strong magnetic field responsible for Larmor precession. 

On the one hand, the added noise amplifies the Larmor-induced peaks to be clearly visible within the power spectral density above the level of detection noise. As a result, our model describing the spin noise in the device can now be used by signal-processing tools interpret the detection data better---e.g., when operating the device as a scalar magnetometer, Bayesian inference methods such as the Kalman filter~\cite{Jimenez2018} can be applied to track fast changes of the strong magnetic field in real time beyond the magnetometer bandwidth~\cite{Wilson2020,Li2020}. On the other hand, the induced noise can be used to naturally perform sensing tasks in the so-called \emph{covert} manner~\cite{Hao2022}, i.e.~so that an adversary having access to the output of the magnetometer would not be able to recover the signal without possessing the pre-calibrated dynamical model that we propose. In this sense, we expect our model to be useful also for tracking time-varying signals encoded in oscillating RF-fields (amplitude/phase) directed perpendicularly to both the scalar field and the added noise.

It would be interesting to generalise our model to a device that operates at the quantum limit~\cite{Zapasskii2013,Glasenapp2014,Shah2010}, i.e.~with detection noise being small enough, so that the predicted peaks in the spectrum arise without the need to artificially amplify them by applying the excess classical noise. Moreover, similarly to orientation-based magnetometers~\cite{lucivero2016squeezed,Troullinou2021,Lucivero2017}, maybe such model could be also capable to incorporate the effect of pumping and probing the ensemble with squeezed light, so the detection noise can be even further reduced. We leave such open questions for the future.

Experimental data created during this research are openly available from the University of Nottingham data repository at \href{URL}{https://doi.org/10.17639/nott.7434}.

\section*{ACKNOWLEDGEMENTS}
This work was supported by  the QuantERA grant C’MON-QSENS! funded by the Engineering and Physical Sciences Research Council (EPSRC) (Grant No.~EP/T027126/1).
Project C’MON-QSENS! is also supported by the National Science Centre (2019/32/Z/ST2/00026), Poland under QuantERA, which has received funding from the European Union's Horizon 2020 research and innovation programme under grant agreement no.~731473. The work was also supported by the Novo Nordisk Foundation (Grant No.~NNF20OC0064182) and the UK Quantum Technology Hub in Sensing and Timing, funded EPSRC (Grant No.~EP/T001046/1) and EPSRC Grant No.~EP/Y005260/1.

\appendix

\section{Evolution in the basis of spherical tensor operators} 
\label{app:spherical_tensors}
The evolution of the atomic ensemble without the effect of pumping, described by \eqnref{eq:rho_evolution_stoch}, is written in terms of the density matrix \eqref{eq:rho_FM}. In this appendix we show how to write it in terms of the vector of all spherical tensor components $\mvec m_t^{\mrm{full}}$. In particular, since the right-hand side of the equation is linear in terms of $\rho$, we would like to find specific linear operators that act on $\mvec m_t^{\mrm{full}}$ that correspond to specific operations performed on $\rho$, and enable us to write the dynamical equation in the form of \eqref{eq:m_evolution1}. We also show that, since the resulting operators do not couple spherical tensor components of different rank, we only need the rank-2 component for a full description of the problem under consideration.
\subsection{Rotations induced by magnetic fields}
Spherical-tensor operators, $\Ttens^{(\kappa)}_q$, are defined by the decomposition of the density matrix \eqref{eq:rho_FM} with respect to irreducible representations of the 3D-rotation group, i.e for each $\kappa=0,..,2F$, operators $\Ttens^{(\kappa)}_q$ with $q=-\kappa,...,\kappa$ form a basis of a $(2\kappa+1)$-dimensional irreducible representation. These operators form a convenient basis for density matrices of a system with fixed angular momentum $F$:
\begin{equation}
    {\rho}=\sum_{\kappa=0}^{2F}\sum_{q=-\kappa}^{\kappa}m_{\kappa q}\Ttens^{(\kappa)}_q,
\end{equation}
with a scalar product $\trace\Big[ {\Ttens^{(\kappa)}_q}^\dag\Ttens^{(\kappa')}_{q'}\Big]=\delta_{\kappa\kappa'}\delta_{qq'}$, and with well-defined behaviour under rotations generated by the angular-momentum operators, i.e.~a vector $\hat{\bm{F}}:=(\hat{F}_x,\hat{F}_y,\hat{F}_z)^T$. In particular, any $SO(3)$ rotation can be parameterised either by the axis (represented by a normalised vector $\vec{n}$), and angle of rotation $\varphi$:
\begin{equation}
    \hat{\mcl R}(\vec n,\varphi)=\ee^{-\ii \varphi \vec{n} \cdot \hat{\bm{F}}},
\end{equation}
or by the three Euler angles $\phi$, $\theta$, $\psi$ that correspond to subsequent rotations about the $z$-, $y$- and again the $z$-axes, respectively:
\begin{equation}
    \hat{\mcl R}(\psi,\theta,\phi)=\ee^{-\ii \psi \hat F_z}\ee^{-\ii \theta \hat F_y}\ee^{-\ii \phi \hat F_z}.
\end{equation}
We now use the operator above to define the Wigner D-matrix:
\begin{align}
    D^{(F)}_{M,M'}(\psi,\theta,\phi) :=\langle F,M|\hat{\mcl R}(\psi,\theta,\phi)|F,M' \rangle.
\end{align}
The spherical tensor operators behave analogously to angular momentum eigenstates under rotations if we treat the rank $\kappa$ as the total angular momentum, and the index $q$ as the projection on the $z$ axis:
\begin{equation}
    \hat{\mcl R}(\psi,\theta,\phi) \Ttens^{(\kappa)}_{q'} \hat{\mcl R}(\psi,\theta,\phi)^\dag =\sum_{q=-\kappa}^\kappa D^{(\kappa)}_{q,q'}(\psi,\theta,\phi) \Ttens^{(\kappa)}_{q}.
\end{equation}
This means that for the spherical tensor component vector $\mvec m_t^{(\kappa)}$ composed of $m_{\kappa,q}(t)$ coefficients ($q=-\kappa,...,\kappa$), the rotation generators are $\mmat{J}^{(\kappa)}_x$, $\mmat{J}^{(\kappa)}_y$ and $\mmat{J}^{(\kappa)}_z$, which are the matrix representations of angular momentum operators cut to the subspace of total angular momentum $\kappa$.

This simplifies the study of dynamics of $\rho$ under rotations, but also facilitates the description of light-atom interactions.
Whenever one considers dipole-type interactions, which correspond to multiplying two (dipole) vectors, spherical-tensor components with $\kappa\le2$ describing the atom are sufficient to find the output light state.~\cite{Happer1972,Bevilacqua2014}. 

We use these properties of the spherical tensor to find the expressions for the evolution of the atomic state described by \eqnref{eq:rho_evolution_stoch} in the spherical tensor basis. Since the Hamiltonian resulting from the magnetic field generates the rotation of the atomic state about the magnetic field vector $\vec B$:
\begin{equation}
    \frac{\dd\rho}{\dd t}=-\ii [\rmdn{\gam}{gmr}\vec B \cdot \hat{\vec F},\rho] ,
\end{equation}
the evolution in the spherical tensor basis will also be driven by respective rotation generators:
\begin{equation}
    \frac{\dd \mvec m_t^{(\kappa)}}{\dd t} = -\ii \rmdn{\gam}{gmr} (\vec B \cdot \vec{J}^{(\kappa)}) \mvec m_t^{(\kappa)},
\end{equation}
where $\vec{J}^{(\kappa)}=\left(\mmat{J}^{(\kappa)}_x, \mmat{J}^{(\kappa)}_y, \mmat{J}^{(\kappa)}_z\right)^\TT$.

This can be directly shown using the commutation relations for the spherical tensor operators and the angular momentum operators:
\begin{align}
    [\hat{F}_z,\Ttens^{(\kappa)}_q ]&=\hbar q \Ttens^{(\kappa)}_q, \label{eq:app_commutation} \\
    [\hat{F}_\pm,\Ttens^{(\kappa)}_q ]&=\hbar \sqrt{\kappa(\kappa+1)-q(q\pm1 )} \Ttens^{(\kappa)}_{\pm q}, \label{app:commutator_Fp_Fm}
\end{align}
that enable us to find the exact form of the matrix:
\begin{align}
    (\mmat A_0)^{(\kappa),q}_{(\kappa'),q'}&=-\frac{\ii \Omega_L}{\hbar}\tr \{ \Ttens^{(\kappa)\dag}_{q} [\hat F_z,\Ttens^{(\kappa')}_{q'}] \} \nonumber \\
    &=-\ii \Omega_L q\delta_{\kappa \kappa'}\delta_{q q'}= -\ii \Omega_L (\mmat{J}^{(\kappa)}_z)^q_{q'}. 
\end{align}
Using \eqnref{app:commutator_Fp_Fm} we could analogously find that commuting the density matrix with $\hat F_x$ and $\hat F_y$ is equivalent to acting the operators $\mmat{J}^{(\kappa)}_x$ and $\mmat{J}^{(\kappa)}_y$, respectively, on $\mvec m_t^{(\kappa)}$. We use this to also obtain:
\begin{equation}
    \rmdn{\mmat A}{\noise}=-\ii \sqrt{\omegarf} \mmat{J}^{(2)}_x.
\end{equation}

Let us now note that commuting density matrix twice with some operator:
\begin{equation}
    [\hat F_\alpha,[\hat F_\alpha,\rho]]=2\hat F_\alpha \rho \hat F_\alpha-\{\hat F_\alpha^2,\rho\} \label{app:double_commutator}
\end{equation}
is then equivalent to acting with the square of the respective operator $\mmat{J}^{(\kappa)}_\alpha$. This is responsible for the appearance of the $\rmdn{\mmat A}{\noise}^2/2$ term in \eqnref{eq:m_evolution1}. In either case, the operation does not couple spherical tensor coefficients of different rank.

\subsection{Decoherence}

We would also like to find the correct description of decoherence using the spherical tensor coefficients. It is driven by non-unitary evolution, described by an operator: 
\begin{equation}
   \frac{\dd\rho}{\dd t}=\Phi[\rho]=\sum_\alpha\Phi_\alpha[\rho]+\Phi_{\mrm{iso}}[\rho],
\end{equation}
where $\Phi_\alpha[\rho]$ is the part of dissipation that comes form the unknown fluctuations of the magnetic field:
\begin{align}
    \Phi_\alpha[\rho]&=\GG_\alpha (\hat F_\alpha \rho \hat F_\alpha-\frac 1 2 \{ \hat F_\alpha^2, \rho \} ), \label{app:Phi_alpha}
\end{align}
and the isotropic part of the dissipation reads: 
\begin{equation}
    \Phi_{\mrm{iso}}[\rho]=\Lambda-\frac 1 2\{\hat\GG,\rho \},
    \label{eq:Phi_iso_app}
\end{equation}
where $\hat \GG=\rmdn{\GG}{iso} \id_{2F+1}$ and $\Lambda=\frac{\rmdn{\GG}{iso}}{2F+1} \, \id_{2F+1}$. It causes the decay of all matrix components at the same rate $\rmdn{\GG}{iso}$ ($-\frac 1 2\{\hat\GG,\rho \}$) and pumps the $m_{0,0}$ components in the rate that balances the decay ($\Lambda$). This means that all of the $\kappa\neq0$ components of the spherical tensor decay at the same rate $\rmdn{\GG}{iso}$:
\begin{equation}
    \frac{\dd m_{\kappa,q}}{\dd t}=-\rmdn{\GG}{iso}m_{\kappa,q}.
\end{equation}
The operator \eqref{app:Phi_alpha}, on the other hand, is proportional to a double commutator of \eqnref{app:double_commutator}, therefore we obtain the following operator of decoherence in the spherical-tensor basis:
\begin{align}
    \mmat A_\Phi=-\frac 1 2 \sum_\alpha \GG_\alpha (\mmat{J}^{(2)}_\alpha)^2-\rmdn{\GG}{iso}\openone_5,
\end{align}
which for the case $\GG_x=\GG_y=\GG_\perp$, $\GG_z=\GG_\parallel$ takes the form:
\begin{equation}
    \mmat{A}_{\Phi}=-\left( \begin{array}{ccccc}
    \GG_2 & 0 & 0 & 0 & 0 \\
    0 & \GG_1 & 0 & 0 & 0 \\
    0 & 0 & \GG_0 & 0 & 0 \\
    0 & 0 & 0 & \GG_1 & 0 \\
    0 & 0 & 0 & 0 & \GG_2
    \end{array} \right),
\end{equation}
with:
\begin{align}
  \GG_0 & =  3 \GG_\perp+\rmdn{\GG}{iso}, \\   
  \GG_1 & =  \frac 1 2 (\GG_\parallel+ 5 \GG_\perp)+\rmdn{\GG}{iso},\\   
  \GG_2 & = 2\GG_\parallel+ \GG_\perp+\rmdn{\GG}{iso}.
\end{align}
We see again that spherical tensor coefficients of different rank do not couple, so we can limit our considerations to the rank-2 component, relevant to our physical system.

\section{Noise spectrum resulting from an inhomogeneous linear SDE} 
\label{app:spectrum_from_equation}
For the purpose of predicting the spin-noise spectrum, we need to consider the following multiple-variable stochastic differential equation:
\begin{equation}
\dd\mvec{m}_t=(\mvec v + \mmat F \mvec{m}_t)\dd t +\mmat G \mvec{m}_t \, \dd W_t, \label{eq:inhomogenous_SDE_MV_app}
\end{equation}
where $\mvec{m}_t$ is an evolving vector, while operators $\mvec F$ and $\mmat G$ together with vector $\mvec v$ parameterise the evolution of the system. The equation is linear and inhomogeneous. Equations like this, in general, are not analytically solvable if the operators $\mvec F$ and $\mmat G$ do not commute. However, we only need to calculate the Fourier transform of the time-autocovariance matrix:
\begin{equation}
\bm \Xi(t)=\langle \mvec{m}_0, \mvec{m}_t^{\mrm T} \rangle_{\mrm{ss}}=\langle \mvec{m}_0 \mvec{m}_t^{\mrm T} \rangle_{\mrm{ss}}-\langle \mvec{m}_0\rangle_{\mrm{ss}} \langle \mvec{m}_t^{\mrm T} \rangle_{\mrm{ss}}. \label{app:Xi_definition}
\end{equation}
In order to obtain the steady-state mean value $\langle \, . \, \rangle_{\mrm{ss}}$, we average over the possible paths of the stochastic process $W_t$ given particular value of the initial point $\mvec{m}_0$, and then separately over all values of the initial point. We indicate these subsequent stochastic averages explicitly by the distinct subscripts, i.e.:
\begin{equation}
\langle \mvec{m}_0 \mvec{m}_t^\TT\rangle_{\mrm{ss}}=\langle \mvec{m}_0 \langle \mvec{m}_t^\TT\rangle_{W_t|\mvec m_0} \rangle_{\mvec{m}_0}.
\end{equation}
Path-averaging of \eqnref{eq:inhomogenous_SDE_MV_app}, which cancels the stochasic increment, leads to
\begin{equation}
\frac{\dd}{\dd t}\langle \mvec{m}_t \rangle_{W_t|\mvec m_0}=\mvec v + \mvec F \langle \mvec{m}_t \rangle_{W_t|\mvec m_0},
\end{equation}
which yields the following evolution of the average value:
\begin{equation}
\langle \mvec{m}_t \rangle_{W_t}=-\mvec F^{-1}\mvec v+\ee^{\mvec F t}\left( \mvec{m}_0+\mvec F^{-1}\mvec v\right).  \label{app:mean_mt_given_m0}
\end{equation}
The steady-state mean of $\mvec m_t$, which is only obtained if all eigenvalues of $\mvec F$ have negative real parts, can be found by taking the limit $t\rightarrow +\infty$:
\begin{equation}
\langle \mvec{m} \rangle_{\mrm{ss}}= -\mvec F^{-1} \mvec v, \label{eq:steady_state_x_mean}
\end{equation}
where we drop the subscript $t$ in $\mvec{m}$, because by definition the steady-state mean value does not evolve. 

Finally, by substituting the expressions \eqref{app:mean_mt_given_m0} and \eqref{eq:steady_state_x_mean} into \eqnref{app:Xi_definition} we obtain:
\begin{equation}
\langle \mvec{m}_0 , \mvec{m}_t^{\mrm T}  \rangle_{\mrm{ss}}=\left[ \langle \mvec{m} \mvec{m}^\TT \rangle_{\mrm{ss}} - \mvec F^{-1}\mvec v \mvec v^\TT (\mvec F^{-1})^\TT \right]\ee^{\mvec F^\TT t},
\end{equation}
which is true for $t\geq 0$. 
An analogous result for $t<0$ can be found by the use of the fact that for the steady state:
\begin{equation}
\langle \mvec{m}_0 , \mvec{m}_{-t}^{\mrm T}  \rangle_{\mrm{ss}} = \langle \mvec{m}_t , \mvec{m}_0^{\mrm T}  \rangle_{\mrm{ss}} = \langle \mvec{m}_0 , \mvec{m}_t^{\mrm T}  \rangle ^\TT,
\end{equation}
so that for any $t$ we can finally write:
\begin{align}
&\langle \mvec{m}_0 , \mvec{m}_t^{\mrm T}  \rangle_{\mrm{ss}} 
= \\
&\quad\left\{\begin{array}{ll}
\left[ \langle \mvec{m} \mvec{m}^\TT \rangle_{\mrm{ss}} - \mvec F^{-1}\mvec v \mvec v^\TT (\mvec F^{-1})^\TT \right]\ee^{\mvec F^\TT t} & \textrm{for } t<0, \\
\ee^{-\mvec F^\TT t} \left[ \langle \mvec{m} \mvec{m}^\TT \rangle_{\mrm{ss}} - \mvec F^{-1}\mvec v \mvec v^\TT (\mvec F^{-1})^\TT \right] & \textrm{for } t\geq 0. \\
\end{array} \right.
\nonumber
\end{align}

In order to get the mean steady-state value of $\langle \mvec{m} \mvec{m}^\TT\rangle$, we need to find
\begin{equation}
\dd  (\mvec{m}_t \mvec{m}_t^\TT) = \dd \mvec{m}_t \, \mvec{m}_t^\TT+ \mvec{m}_t \, \dd \mvec{m}_t^\TT +\dd \mvec{m}_t \, \dd \mvec{m}_t^\TT.
\end{equation}
The $\dd \mvec{m}_t \, \dd \mvec{m}_t^\TT$ term is necessary to account for the property of the stochastic increment that $\dd W_t^2=\dd t$. We substitute $ \dd\mvec m_t $ from \eqnref{eq:inhomogenous_SDE_MV_app}, take the average, divide by $\dd t$ and obtain:
\begin{align}
\frac{\dd}{\dd t} \langle \mvec{m}_t \mvec{m}_t^\TT\rangle &= \mvec v \langle \mvec{m}_t^\TT \rangle + \langle \mvec{m}_t \rangle \mvec v^\TT + \mvec F\langle \mvec{m}_t \mvec{m}_t^\TT \rangle \nonumber \\
&+ \langle \mvec{m}_t \mvec{m}_t^\TT \rangle \mvec F^\TT + \mmat G \langle \mvec{m}_t \mvec{m}_t^\TT \rangle \mmat G^\TT,
\end{align}
which equals 0 in the steady state. Hence, after substituting $\langle\mvec m_t\rangle=\langle\mvec m\rangle_{\mrm{ss}}$ from \eqref{eq:steady_state_x_mean}, we obtain:
\begin{equation}
\mvec F \, \bm \sigma + \bm \sigma \, \mvec F^{\mrm T} + \mmat G \, \bm \sigma \, \mmat G^{\mrm T}=\mvec v \, \mvec v^\TT \, (\mvec F^\TT)^{-1}+\mvec F^{-1} \, \mvec v \, \mvec v^\TT, \label{eq:sigma_equation}
\end{equation}
where $\bm \sigma=\langle \mvec{m} \, \mvec{m}^{\mrm T} \rangle_{\mrm{ss}}$. Importantly, in case we are given a specific set of $\mvec v$, $\mvec F$ and $\mmat G$, we can solve \eqref{eq:sigma_equation} as a system of linear equations to find $\bm \sigma$. 

Possessing a particular form of $\bm \sigma$, we can explicitly calculate the Fourier transform of $\bm\Xi (t)$, i.e.:
\begin{align}
\bm \Xi(\ww)
&= \int_{-\infty}^0\!\!\! \dd t \,\ee^{-\mvec F^\TT t} \left[ \langle \mvec{m} \mvec{m}^\TT \rangle - \mvec F^{-1}\mvec v \mvec v^\TT (\mvec F^{-1})^\TT \right] \\
&+ \int_0^{+\infty} \!\!\! \dd t\left[\bm \sigma-\mvec F^{-1}\mvec v \mvec v^\TT (\mvec F^\TT)^{-1}\right]\exp\left[ (-\ii\ww+\mvec F^\TT)t \right] \nonumber \\
&= -\frac{1}{2\pi} (\mvec F + \ii \ww)^{-1} \left[\bm \sigma-\mvec F^{-1}\mvec v \mvec v^\TT (\mvec F^\TT)^{-1}\right] \\
&-\frac{1}{2\pi} \left[\bm \sigma-\mvec F^{-1}\mvec v \mvec v^\TT (\mvec F^\TT)^{-1}\right](\mvec F^\TT - \ii \ww)^{-1}. \nonumber
\end{align}
We can simplify this formula by using \eqnref{eq:sigma_equation} and finding that:
\begin{equation}
(\mvec F + \ii \ww) \bm \Xi(\ww)
 (\mvec F^\TT - \ii \ww)= \mmat G\, \bm \sigma\, \mmat G^\TT,
\end{equation}
so we obtain:
\begin{equation}
\bm \Xi(\ww) = (\mvec F + \ii \ww)^{-1} \mmat G\, \bm \sigma\, \mmat G^\TT (\mvec F^\TT - \ii \ww)^{-1}.
\end{equation}

If the signal is a linear combination of the $\mvec m_t$ components, parameterised by a vector $\mvec k$ of the same dimension as $\mvec m_t$:
\begin{equation}
   S(t)=\mvec k^\TT\, \mvec x_t,
\end{equation}
then the PSD is given by:
\begin{align}
    \mrm{PSD}(\ww)&=\mvec k^\TT\, \bm \Xi(\ww)\,\mvec k = \nonumber \\
    &=\mvec k^\TT\, (\mvec F + \ii \ww)^{-1} \mmat G\, \bm \sigma\, \mmat G^\TT (\mvec F^\TT - \ii \ww)^{-1}\,\mvec k . \label{app:PSD_omega}
\end{align}
\\

\subsection{Effective Lorentzian form} \label{app:spec_form}
To find the functional form of $\mrm{PSD}(\ww)$ one can write \eqnref{app:PSD_omega} in the eigenbasis of the $\mmat F$ matrix. If we write the eigenvalues of $\mmat F$ as $\lambda_\alpha=-\gam_\alpha+\ii\omega_\alpha$, then 
\begin{align}
    \mrm{PSD}(\ww)=&\sum_{\alpha,\beta}\frac{k_\alpha k_\beta Q_{\alpha\beta}}{[-\gam_\alpha+\ii(\ww+\ww_\alpha)][-\gam_\beta-\ii(\ww-\ww_\beta)]} \nonumber \\
    =&\sum_{\alpha,\beta} k_\alpha k_\beta Q_{\alpha\beta}\frac{\gam_\alpha+\gam_\beta+\ii (\ww_\alpha+\ww_\beta)}{(\gam_\alpha+\gam_\beta)^2+(\ww_\alpha+\ww_\beta)^2} \nonumber \\
    &\times\left(\frac{\gam_\alpha+\ii(\ww+\ww_\alpha)}{\gam_\alpha^2+(\ww+\ww_\alpha)^2}+\frac{\gam_\beta-\ii(\ww-\ww_\beta)}{\gam_\beta^2+(\ww-\ww_\beta)^2}\right),
\end{align}
where $ k_\alpha$, $ k_\beta$ are the components of the $\mvec k$ vector and $ Q_{\alpha \beta}$ are the components of the matrix $\mmat G\, \bm \sigma\, \mmat G^\TT$, in both cases written in the eigenbasis of $\mvec F$. As one can see, the resulting spectrum is a sum of symmetric (absorptive) and antisymmetric (dispersive) Lorentzian peaks, whose widths are the opposites of real parts ($\gamma_\alpha$), and central frequencies are the imaginary parts of the eigenvalues and their opposites ($\pm\ww_{\alpha}$):
\begin{align}
    \mrm{PSD}(\ww)=\sum_\alpha \left( \frac{p_\alpha^a \gam_\alpha^2}{\gam_\alpha^2+(\ww-\ww_\alpha)^2} + \frac{p_\alpha^a \gam_\alpha^2}{\gam_\alpha^2+(\ww+\ww_\alpha)^2}\right. \nonumber \\
    \left. +\frac{p_\alpha^d \gam_\alpha(\ww-\ww_\alpha)}{\gam_\alpha^2+(\ww-\ww_\alpha)^2} + \frac{p_\alpha^d \gam_\alpha(\ww+\ww_\alpha)}{\gam_\alpha^2+(\ww+\ww_\alpha)^2}\right),
\end{align}
which, in the case of a real signal $S(t)$ we know to be real, positive, and symmetric around $\ww=0$.

\section{Analytic prediction of peak amplitudes} 
\label{app:peak_amp}
Crucially, the atomic magnetometer under study, described by dynamics \eref{eq:dx_full}, constitutes an example of stochastic inhomogeneous evolution discussed in \appref{app:spectrum_from_equation}, so that the PSD without external noise $\zeta(t)$ is given by \eqnref{app:PSD_omega} with $\mvec v = \mmat{A}_{\Phi} \mvec{m}^\ss$, $\mmat F = \mmat{A}_0 + \mmat{A}_{\Phi} + \frac{\rmdn{\mmat A}{\noise}^2}{2}$, $\mmat G=\rmdn{\mmat A}{\noise}$ and $\mvec k=\mmat{D}^{(2)\TT}_\theta\,\mvec{h}$.

Since the characteristic polynomial of the matrix $\mmat F$ has real coefficients, we know that it has at least one real eigenvalue, while the other two are, in pairs, complex conjugates of each other. This limits the number and character of peaks to one symmetric peak at zero, and two pairs of peaks of opposite frequencies, that are sums of symmetric and antisymmetric peaks:
\begin{align}
    \mrm{PSD}(\ww)&=\sum_{j=-2,..,2}\frac{p_{|j|}^a\gam_{|j|}^2}{(\ww-\ww_j)^2+\gam_{|j|}^2} \\
    &+ \sum_{j=\pm 1,\pm 2}\frac{\pm p_{|j|}^d\gam_{|j|}(\ww-\ww_j)}{(\ww-\ww_j)^2+\gam_{|j|}^2},
\end{align}
where the approximate $\ww_j$ and $\gamma_j$ are given by \eqnsref{eq:omega_j}{eq:gamma_j} of main text, respectively.

\begin{figure*}[t!]
    \centering
    \subfigure[]{\includegraphics[width=\columnwidth]{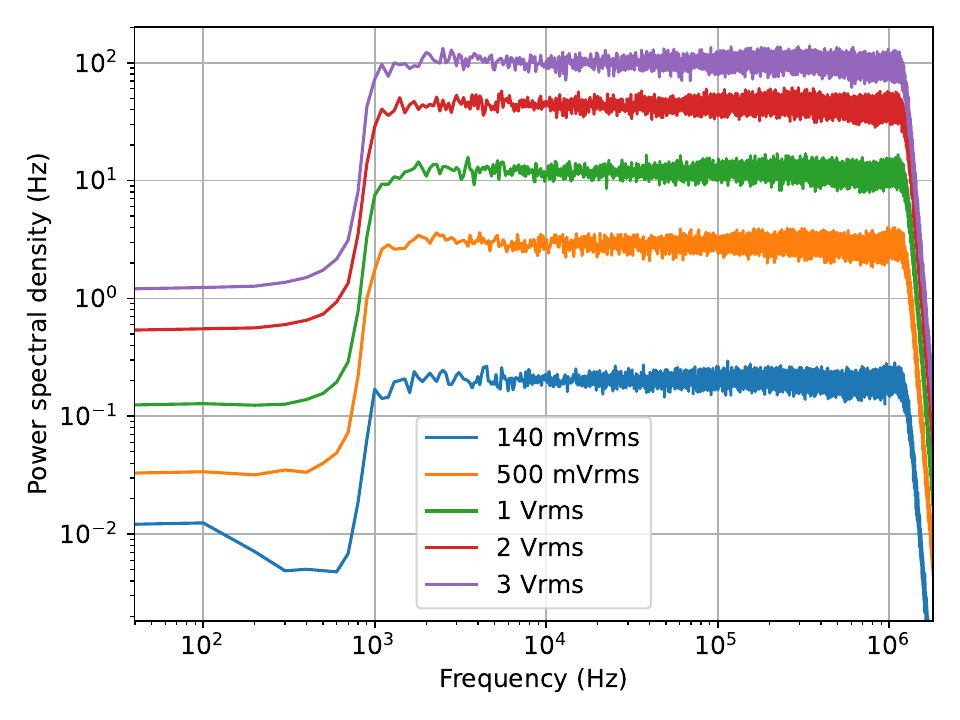}\label{fig:calibration_PSD}}%
    \subfigure[]{\includegraphics[width=\columnwidth]{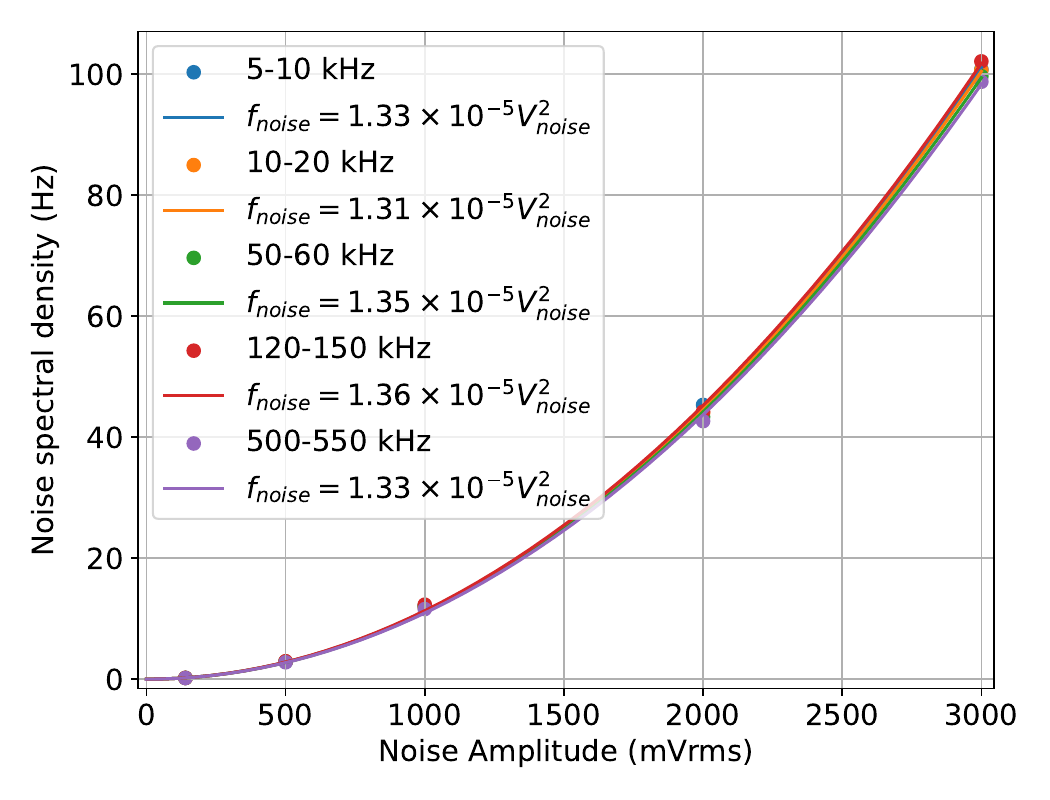}\label{fig:calibration_fit}}    
    \caption{
    \textbf{Amplitude calibration for the noise spectral density} $\omegarf=2\pi\frf$. 
    (a) The output current of the noise generator is first transmitted through both $1$ kHz high-pass and $1$ MHz low-pass filters, before being directly measured.  100 independent time traces of 0.01 s duration (sampled at a rate 40MHz) are recorded to compute the resulting \emph{power spectral density} (PSD) shown, which is converted into the units of Hz after accounting for the calibration factor of the coil inducing the noisy magnetic field and the gyromagnetic ratio of caesium.
    (b) In order to obtain the effective magnitude of the PSD, i.e. the \emph{noise spectral density}, the average value of the PSD is computed for five different frequency ranges, as a function of the amplitude $\Vrf$ of the noise applied. The plot shows quadratic fits, $f_\trm{noise} = c \Vrf^2$, applicable to each of the five frequency ranges used for averaging. All these agree and yield $c = 1.33(3) \times 10^{-5}$~Hz/mV$_{\mathrm{rms}}^2$.
    }
\end{figure*}

The formula resulting from \eqnref{app:PSD_omega} has a complex form, practically only possible to obtain and store using symbolic calculation software, therefore it does not provide a clear picture of the properties of the spectrum, except for making it possible to plot it numerically. In order to find the specific values of $p_i^a$, $p_i^d$, which are hidden in the formula, we use the observation that for $\Omega_L\gg\gam_i$ the heights of the peaks in the spectrum do not depend on $\Omega_L$, and therefore the limits:
\begin{equation}
    p_i^a\approx\lim_{\Omega_L\rightarrow +\infty}\mrm{PSD}(\ww=\ww_i)
\end{equation}
will give, in sufficiently good approximation, the values of $p_i^a$. To obtain the approximate values of $p_i^d$, we calculate:
\begin{equation}
    p_i^d\approx\frac {\GG_i} {\Omega_L}\lim_{\Omega_L\rightarrow +\infty}\left( \Omega_L\left. \frac{\dd}{\dd \ww}\mrm{PSD}(\ww)\right|_{\ww=\ww_i}\right),
\end{equation}
because we need the second order of expansion in $\frac{1}{\Omega_L}$. Using this approximation we obtain Eqs. \eqref{eqn:peaks_omega} from the main text, and:
\begin{widetext}
\begin{align}
    p^d_1=& g_D^2 \frac{3}{2}(m^{\mrm{ini}}_{20})^4 \frac{\omegarf}{\Omega_L} \frac{\GG_0^2(16\GG_1^2\GG_2+8\GG_1(\GG_1+5\GG_2)\omegarf+16(\GG_1+\GG_2)\omegarf^2+3\omegarf^3)}{(4\GG_1+5 \omegarf)^3\,G(\omegarf,\vec{\GG})} [3+2\cos(2\theta)+3\cos(4\theta)]^2 \label{app:peak_d1_height}\\
    p^d_2=& g_D^2 \frac{9}{64}(m^{\mrm{ini}}_{20})^4\frac{\omegarf^2}{\Omega_L}\frac{\GG_0^2(8\Gamma_2^2+8\GG_2 \omegarf+\omegarf^2)}{(2\GG_2+ \omegarf)^3\,G(\omegarf,\vec{\GG})} [2\sin(2\theta)+3\sin(4\theta)]^2 \label{app:peak_d2_height}
\end{align}
\end{widetext}
where $G(\omegarf,\vec{\GG})$ is stated in \eqnref{eq:G_function} of the main text.

\subsection{Irrelevance of the dispersive contributions} 
\label{app:disp_contr}
We note that we have not verified the predicted relations for the dispersive contributions of the line shape, see \eqnref{app:peak_d1_height} and \eqnref{app:peak_d2_height} in App.~\ref{app:peak_amp}. This is due to their contribution being below the noise floor of our system. We can verify that this is as expected for a Larmor frequency of approximately 9.45~kHz by calculating the ratios of the peak height equations and using the linewidths found in the experiment: $\Gamma_1 = 34$~Hz and $\Gamma_2 = 34$~Hz. We find $p^d_1 / p^a_1$ from Eq.~(\ref{app:peak_d1_height}) divided by Eq.~(\ref{eqn:peaks_omega1}) and we find $p^d_2 / p^a_2$ from Eq.~(\ref{app:peak_d2_height}) divided by Eq.~(\ref{eqn:peaks_omega2}). 
The calculations were done for the lowest white noise amplitude of $7~\mathrm{mV_{rms}}$ and the highest white noise amplitude of $3.5~\mathrm{V_{rms}}$. For $\mathrm{V_{noise}} = 7~\mathrm{mV_{rms}}$ we find $p^d_1 / p^a_1 = 1.1 \times 10^{-4}$ and $p^d_2 / p^a_2 = 3.2 \times 10^{-4}$. Furthermore, when $\mathrm{V_{noise}} = 3.5~\mathrm{V_{rms}}$ we find $p^d_1 / p^a_1 = 3.1 \times 10^{-5}$ and $p^d_2 / p^a_2 = 1.8 \times 10^{-4}$. Hence, the dispersive contribution is at least 4 orders of magnitude smaller than the absorptive contribution. Hence, this is significantly below our noise floor for both peaks, as seen experimentally.

\section{Amplitude calibration for the noise spectral density $\omegarf=2\pi\frf$}
\label{app:Calibration}
In order to determine the value of the constant $c$ relating the magnitude of the noise spectral density to the (RMS) amplitude of the voltage applied in the noise generator, i.e.~$\omegarf = 2 \pi \frf = 2 \pi c \Vrf^2$, we directly measure the current in the coil that induces the noisy magnetic field. The signal from the generator, however, is first transformed thorough a 1~MHz low-pass filter to prevent the effect of aliasing, as well as a 1~kHz high-pass filter to eliminate any spurious contributions at very low (DC-like) frequencies. We record 100 traces of 0.01~s duration of the current reaching the coil (sampled at 40 MHz), while varying the voltage amplitude in the noise generator. Based on these we compute the PSD \eref{eq:PSD}, which is presented in \figref{fig:calibration_PSD} in the units of $\mathrm{Hz}^2/\mathrm{Hz}$, after performing adequate rescaling given the values of the coil calibration factor (10.1~$\mathrm{nT/mV_{rms}}$) and the gyromagnetic ratio of caesium (3.5~Hz/nT).

From \figref{fig:calibration_PSD} we can determine  $\frf$ by seeing how the PSD varies with different white-noise amplitudes. We average areas of \figref{fig:calibration_PSD} in five different frequency ranges (specified in the label of \figref{fig:calibration_fit}) to obtain the effective value of $\frf$ as a function of the noise amplitude---five values of $\Vrf$ (stated in $\mathrm{mV_{rms}}$) are used. For each of the averaged ranges, we then fit the quadratic dependence, $\frf=c V_{\mathrm{\mrm{\noise}}}^2 $, what allow us to overall obtain $c = 1.33(3) \times 10^{-5}$~Hz/mV$_{\mathrm{rms}}^2$, see \figref{fig:calibration_fit}.

\bibliographystyle{myapsrev4-2}
\bibliography{align_SNS}

\end{document}